\documentclass[preprint,eqsecnum,aps,nofootinbib]{revtex4}
\usepackage{amsfonts,amsmath,amssymb,amsthm}
\usepackage{latexsym}
\usepackage{bbm,bm}
\usepackage{graphicx}

\newcommand{\ket}[1]{\lvert #1 \rangle}
\newcommand{\bra}[1]{\langle #1 \lvert}
\newcommand{\beq}{\begin{equation}}
\newcommand{\eeq}{\end{equation}}
\newcommand{\beqs}{\begin{eqnarray}}
\newcommand{\eeqs}{\end{eqnarray}}

\begin{document}

\title{Testing the Monogamy Relations via Rank-2 Mixtures}

\author{Eylee Jung$^{1}$ and DaeKil Park$^{1,2}$}

\affiliation{$^1$Department of Electronic Engineering, Kyungnam University, Changwon
                 631-701, Korea    \\
             $^2$Department of Physics, Kyungnam University, Changwon
                  631-701, Korea    
                      }

\begin{abstract}
We introduce two tangle-based four-party entanglement measures $t_1$ and $t_2$, and two negativity-based measures $n_1$ and $n_2$, which are derived 
from the monogamy relations. These measures are computed for three four-qubit maximally entangled and W states explicitly. We also compute these 
measures for the rank-$2$ mixture $\rho_4 = p \ket{\mbox{GHZ}_4} \bra{\mbox{GHZ}_4} + (1 - p) \ket{\mbox{W}_4} \bra{\mbox{W}_4}$ by finding the 
corresponding optimal decompositions. It turns out that $t_1 (\rho_4)$ is trivial and the corresponding optimal decomposition is equal to the 
spectral decomposition. Probably, this triviality is a sign of the fact that the corresponding monogamy inequality is not sufficiently tight. We fail to compute 
$t_2 (\rho_4)$ due to the difficulty for the calculation of the residual entanglement. The negativity-based measures $n_1 (\rho_4)$ and $n_2 (\rho_4)$ are 
explicitly computed and the corresponding optimal decompositions are also derived explicitly.
\end{abstract}

\maketitle

\section{Introduction}
Research into entanglement of quantum states has long history from the very beginning of quantum mechanics\cite{epr-35,schrodinger-35}. At 
that time the main motivation for the study of entanglement was pure theoretical. It was to explore the non-local properties of quantum mechanics. 
Recent considerable attention to the quantum entanglement\cite{text,horodecki09} has both theoretical and practical aspects. While the former is for 
understanding of quantum information theories more deeply, the latter is for developing the quantum technology.
As shown for last two decades quantum entanglement plays a central role in quantum teleportation\cite{teleportation},
superdense coding\cite{superdense}, quantum cloning\cite{clon}, and quantum cryptography\cite{cryptography,cryptography2}. It is also quantum entanglement, which makes the quantum computer\footnote{The current status of quantum computer technology was reviewed in Ref.\cite{qcreview}.} outperform the classical one\cite{computer}. Thus, it is very important to understand how to quantify 
and how to characterize the entanglement. Still, however, this issue is not completely understood.

For bipartite quantum system many entanglement measures were constructed before such as distillable entanglement\cite{benn96}, entanglement 
of formation (EOF)\cite{benn96}, and relative entropy of entanglement (REE)\cite{vedral-97-1,vedral-97-2}. Among them\footnote{Although there are 
a lot of attempts to derive the closed formula for REE, still we do not know how to compute the REE for the arbitrary two qubit mixtures except rare cases
\cite{ree}.} the closed formula for the analytic computation of EOF for states of two qubits were found in Ref. \cite{woot-98} via the concurrence ${\cal C}$ as 
\begin{equation}
\label{EoF-1}
E_F ({\cal C}) = h \left( \frac{1 + \sqrt{1 - {\cal C}^2}}{2} \right),
\end{equation}
where $h(x)$ is a binary entropy function $h(x) = -x \ln x - (1 - x) \ln (1 - x)$. For two-qubit pure state 
$\ket{\psi}_{AB} = \psi_{ij} \ket{ij}_{AB}$ with $(i,j=0,1)$, the concurrence ${\cal C}_{A|B}$ between party $A$ and party $B$  is given by
\begin{equation}
\label{concurrence-1}
{\cal C}_{A|B} = |\epsilon_{i_1 i_2} \epsilon_{j_1 j_2} \psi_{i_1 j_1} \psi_{i_2 j_2}| = 2 |\psi_{00} \psi_{11} - \psi_{01} \psi_{10}|,
\end{equation}
where the Einstein convention is understood and $\epsilon_{\mu \nu}$ is an antisymmetric tensor. For two-qubit 
mixed state $\rho_{AB}$ the concurrence ${\cal C}_{A|B}(\rho)$ can be computed by ${\cal C}_{A|B} = \max(\lambda_1 - \lambda_2 - \lambda_3 - \lambda_4, 0)$, where $\{\lambda_1^2, \lambda_2^2, \lambda_3^2, \lambda_4^2\}$ are eigenvalues of positive operator
$\rho (\sigma_y \otimes \sigma_y) \rho^* (\sigma_y \otimes \sigma_y)$ with decreasing order. Thus, one can compute 
the EOF for all two-qubit states in principle.

Generalization to the multipartite entanglement is highly important and challenging issue in the context of quantum information theories. A seminal step toward 
this goal was initiated in Ref. \cite{ckw} by examining the three-qubit pure states. Authors in  Ref. \cite{ckw} have shown analytically the monogamy relation
\begin{equation}
\label{ckw}
{\cal C}^2_{q_1|(q_2q_3)} \geq {\cal C}^2_{q_1|q_2} +  {\cal C}^2_{q_1|q_3}.
\end{equation}
This relation implies that the entanglement (measured by the squared concurrence) between $q_1$ and the remaining parties always exceeds 
entanglement between $q_1$ and $q_2$ plus entanglement between $q_1$ and $q_3$. This means that if $q_1$ and $q_2$ is maximally entangled, 
the whole system cannot have the tripartite entanglement. The inequality (\ref{ckw}) is strong in a sense that the three-qubit W-state\cite{dur00}
\begin{equation}
\label{w3}
\ket{\mbox{W}_3} = \frac{1}{\sqrt{3}} \left( \ket{001} + \ket{010} + \ket{100} \right)
\end{equation}
saturates the inequality. Moreover, for three-qubit pure state $\ket{\psi}_{ABC} = \psi_{ijk} \ket{ijk}_{ABC}$ the leftover in the inequality
\begin{equation}
\label{residual}
\tau_{A|B|C} = {\cal C}^2_{A|(BC)} - \left(  {\cal C}^2_{A|B} +  {\cal C}^2_{A|C} \right),
\end{equation}
which we will call the residual entanglement\footnote{In this paper $\sqrt{\tau_{A|B|C}}$ is called the three-tangle.}, has  following two expressions:
\begin{eqnarray}
\label{tangle3}
\tau_{A|B|C} = \bigg|2 \epsilon_{i_1 i_2} \epsilon_{i_3 i_4} \epsilon_{j_1 j_2} \epsilon_{j_3 j_4} \epsilon_{k_1 k_3} \epsilon_{k_2 k_4}
          \psi_{i_1 j_1 k_1} \psi_{i_2 j_2 k_2} \psi_{i_3 j_3 k_3} \psi_{i_4 j_4 k_4} \bigg|   = 4 |d_1 - 2 d_2 + 4 d_3|
 \end{eqnarray}
where 
\begin{eqnarray}
\label{3-tangle-2}
& &d_1 = \psi^2_{000} \psi^2_{111} + \psi^2_{001} \psi^2_{110} + \psi^2_{010} \psi^2_{101} +
                                                              \psi^2_{100} \psi^2_{011
},
                                                              \\   \nonumber
& &d_2 = \psi_{000} \psi_{111} \psi_{011} \psi_{100} + \psi_{000} \psi_{111} \psi_{101} \psi_{010} +
         \psi_{000} \psi_{111} \psi_{110} \psi_{001}
                                                              \\   \nonumber
& &\hspace{1.0cm} +
         \psi_{011} \psi_{100} \psi_{101} \psi_{010} + \psi_{011} \psi_{100} \psi_{110} \psi_{001} +
         \psi_{101} \psi_{010} \psi_{110} \psi_{001},
                                                              \\   \nonumber
& &d_3 = \psi_{000} \psi_{110} \psi_{101} \psi_{011} + \psi_{111} \psi_{001} \psi_{010} \psi_{100}.
\end{eqnarray}
From first expression one can show that $\tau_{A|B|C}$ is invariant under  a stochastic local operation and classical 
communication (SLOCC)\cite{bennet00}. From second expression one can show that $\tau_{A|B|C}$ is invariant under the qubit permutation. It was 
also shown in Ref. \cite{ckw} that $\tau_{A|B|C}$ is an entanglement monotone. Thus, the residual entanglement (or three-tangle) can play a role
as an important measure for the genuine three-way entanglement.

By making use of Eq. (\ref{tangle3}) one can compute the residual entanglement of all three-qubit pure states. For mixed state the residual entanglement 
is usually defined  as a convex roof method\cite{benn96,uhlmann99-1}
\begin{equation}
\label{roof}
\tau_{A|B|C} (\rho) = \min \sum_i p_i \tau_{A|B|C} (\ket{\psi_i} \bra{\psi_i})
\end{equation}
where the minimum is taken over all possible ensembles of pure states. The ensemble corresponding to the minimum of $\tau_{A|B|C}$ is called 
optimal decomposition. For given three-qubit mixed state it is highly difficult, in general, to find its optimal decomposition except very 
rare cases\cite{tangle}\footnote{Recently, the three-tangle of the GHZ-symmetric states\cite{elts12-1} has been computed analytically\cite{siewert12-1}.}.

In order to find the entanglement measures in the multipartite system, there are two different approaches. First approach is to find the invariant 
monotones under the SLOCC transformation. As Ref.\cite{verst03} has shown, any 
linearly homogeneous positive function of a pure state that is invariant under determinant $1$ SLOCC operations is an entanglement 
monotone. Thus, the concurrence ${\cal C}_{A|B}$ and the three-tangle $\sqrt{\tau_{A|B|C}}$ are monotones.
It is also possible to construct the SLOCC-invariant monotones in the higher-qubit systems. In the higher-qubit systems, however, 
there are many independent monotones, because the number of independent SLOCC-invariant monotones is equal to the degrees of freedom
of pure quantum state minus the degrees of freedom induced by the determinant $1$ SLOCC operations. For example, there are 
$2(2^n - 1) - 6 n$ independent monotones in $n$-qubit system. Thus, in four-qubit system there are six invariant monotones. Among them, it was shown in 
Ref. \cite{four-way} by making use of the antilinearity\cite{uhlmann99-1} that there are following three independent monotones which measure the true four-way entanglement:
\begin{eqnarray}
\label{four-measure}
& &{\cal F}^{(4)}_1 = (\sigma_{\mu} \sigma_{\nu} \sigma_2 \sigma_2) \bullet (\sigma^{\mu} \sigma_2 \sigma_{\lambda} \sigma_2) \bullet
                      (\sigma_2 \sigma^{\nu} \sigma^{\lambda} \sigma_2)           \nonumber  \\
& &{\cal F}^{(4)}_2 = (\sigma_{\mu} \sigma_{\nu} \sigma_2 \sigma_2) \bullet (\sigma^{\mu} \sigma_2 \sigma_{\lambda} \sigma_2) \bullet (\sigma_2 \sigma^{\nu} \sigma_2 \sigma_{\tau}) \bullet (\sigma_2 \sigma_2 \sigma^{\lambda} \sigma^{\tau})                                                                 \\    \nonumber
& &{\cal F}^{(4)}_3 = \frac{1}{2} (\sigma_{\mu} \sigma_{\nu} \sigma_2 \sigma_2) \bullet (\sigma^{\mu} \sigma^{\nu} \sigma_2 \sigma_2) \bullet (\sigma_{\rho} \sigma_2 \sigma_{\tau} \sigma_2) \bullet
(\sigma^{\rho} \sigma_2 \sigma^{\tau} \sigma_2) \bullet (\sigma_{\kappa} \sigma_2 \sigma_2 \sigma_{\lambda})
\bullet (\sigma^{\kappa} \sigma_2 \sigma_2 \sigma^{\lambda}),
\end{eqnarray}
where $\sigma_0 = \openone_2$, $\sigma_1 = \sigma_x$, $\sigma_2 = \sigma_y$, $\sigma_3 = \sigma_z$, and
the Einstein convention is introduced with a metric $g^{\mu \nu} = \mbox{diag} \{-1, 1, 0, 1\}$. 
The solid dot in Eq. (\ref{four-measure}) is defined as follows. Let $\ket{\psi}$ be a four-qubit state. Then, for example, ${\cal F}^{(4)}_1$ of $\ket{\psi}$ is 
defined as 
\begin{equation}
\label{revise1}
{\cal F}^{(4)}_1 (\psi) = \bigg| \bra{\psi^*} \sigma_{\mu} \otimes \sigma_{\nu} \otimes \sigma_2 \otimes \sigma_2 \ket{\psi} \bra{\psi^*} \sigma^{\mu} \otimes \sigma_2 \otimes \sigma_{\lambda} \otimes \sigma_2 \ket{\psi}
                                                  \bra{\psi^*} \sigma_2 \otimes \sigma^{\nu} \otimes \sigma^{\lambda} \otimes \sigma_2  \ket{\psi}   \bigg|.
\end{equation}
Other measures can be computed similarly.
Furthermore, 
it was shown in Ref. \cite{oster06-1} that there are following three maximally entangled states in 
four-qubit system:
\begin{eqnarray}
\label{four-maximal}
& &\ket{\mbox{GHZ}_4} = \frac{1}{\sqrt{2}} \big(\ket{0000} + \ket{1111} \big)       \nonumber  \\
& &\ket{\Phi_2} = \frac{1}{\sqrt{6}} \left(\sqrt{2} \ket{1111} + \ket{1000} + \ket{0100} + \ket{0010} + \ket{0001}
                                      \right)                                     \\    \nonumber
& &\ket{\Phi_3} = \frac{1}{2} \big(\ket{1111} + \ket{1100} + \ket{0010} + \ket{0001} \big).
\end{eqnarray}
The measures ${\cal F}^{(4)}_1$, ${\cal F}^{(4)}_2$, and ${\cal F}^{(4)}_3$ of $\ket{\mbox{GHZ}_4}$, $\ket{\Phi_2}$, 
$\ket{\Phi_3}$, and 
\begin{equation}
\label{wtilde-4}
\ket{\tilde{\mbox{W}}_4} = \frac{1}{2} \big( \ket{0111} + \ket{1011} + \ket{1101} + \ket{1110} \big)
\end{equation}
are summarized in Table I. Recently, ${\cal F}^{(4)}_j \hspace{.2cm} (j = 1, 2, 3)$ and the corresponding linear monotones\footnote{The linear monotone means a monotone of homogeneous degree $2$.
Thus, if  ${\cal F}^{(4)}_j$ is a measure of homogeneous degree $D$, the corresponding one is  ${\cal G}^{(4)}_j = \left({\cal F}^{(4)}_j \right)^{2 / D}$.}
${\cal G}^{(4)}_j \hspace{.2cm} (j = 1, 2, 3)$ for the rank-$2$ mixtures consist of one of  the maximally entangled state and $\ket{\tilde{\mbox{W}}_4}$ are explicitly computed\cite{eylee15-1}.

\begin{center}
\begin{tabular}{c|ccc} \hline \hline
& $\hspace{.2cm}{\cal F}^{(4)}_1 \hspace{.2cm}$ &  $\hspace{.2cm} {\cal F}^{(4)}_2 \hspace{.2cm}$  &  
$\hspace{.2cm} {\cal F}^{(4)}_3 \hspace{.2cm}$  \\  \hline 
$\ket{\Phi_1} \hspace{.2cm}$ & $1$ & $1$ & $\frac{1}{2}$     \\   
$\ket{\Phi_2} \hspace{.2cm}$ & $\frac{8}{9}$ & $0$ & $0$      \\  
$\ket{\Phi_3} \hspace{.2cm}$ & $0$ & $0$ & $1$                 \\  
$\ket{\tilde{\mbox{W}}_4} \hspace{.2cm}$  & $0$  &  $0$  &  $0$                \\   \hline  \hline
\end{tabular}

\vspace{0.2cm}
Table I:${\cal F}^{(4)}_1$, ${\cal F}^{(4)}_2$, and ${\cal F}^{(4)}_3$ of the maximally entangled and 
$\tilde{\mbox{W}}_4$ states.
\end{center}

Second approach is to find the monogamy relations in the multipartite system. As Ref. \cite{osborne06-1} has shown analytically  the following monogamy relation
\begin{equation}
\label{tofv}
{\cal C}^2_{q_1|(q_2 \cdots q_n)} \geq {\cal C}^2_{q_1|q_2} +  {\cal C}^2_{q_1|q_3} + \cdots +  {\cal C}^2_{q_1|q_n}
\end{equation}
holds in the $n$-qubit pure-state system. However, the leftover of Eq. (\ref{tofv}) is not entanglement monotone. The authors in Ref. \cite{bai07-1,bai08-1} conjectured
that in four-qubit system the following quantity
\begin{equation}
\label{tangle-1}
t_1 = \frac{\pi_A + \pi_B + \pi_C + \pi_D}{4}
\end{equation}
is a monotone, where $\pi_A = {\cal C}^2_{A|(BCD)} - ({\cal C}^2_{A|B} + {\cal C}^2_{A|C} + {\cal C}^2_{A|D})$ and other ones are obtained by 
changing the focusing qubit. Even though $t_1$ might be an entanglement monotone, it is obvious that it is not a true four-way measure because it 
detects the three-way entanglement. For example, $t_1 (g_3) = 3 / 4$, where $\ket{g_3} = (\ket{0000} + \ket{1110}) / \sqrt{2}$.

In Ref. \cite{regula14-1} another following multipartite monogamy relation is derived:
\begin{equation}
\label{smonogamy-1}
{\cal C}^2_{q_1|(q_2 \cdots q_n)} \geq \underbrace{\sum_{j=2}^n {\cal C}^2_{q_1|q_j}}_{2-\mbox{partite}} + \underbrace{\sum_{k > j=2}^n \left[ \tau_{q_1|q_j|q_k} \right]^{\mu_3}}_{3-\mbox{partite}} + \cdots + 
\underbrace{\sum_{\ell = 2}^n \left[ \tau_{q_1|q_2|\cdots | q_{\ell-1} | q_{\ell+ 1} | \cdots |q_n} \right]^{\mu_{n-1}}}_{(n-1)-\mbox{partite}}.
\end{equation}
In Eq. (\ref{smonogamy-1}) the power factors $\left\{\mu_m \right\}_{m=3}^{n-1}$ are included to regulate the weight assigned to the different
$m$-partite contributions. If all power factors $\mu_m$ go to infinity,   Eq. (\ref{smonogamy-1}) reduces to Eq. (\ref{tofv}). Especially, the authors in Ref. \cite{regula14-1} have conjectured $\mu_3 = 3 / 2$.
Thus, in four-qubit system one can construct another possible candidate of the tangle-based entanglement measure
\begin{equation}
\label{tangle-2}
t_2 = \frac{\sigma_A + \sigma_B + \sigma_C + \sigma_D}{4},
\end{equation}
where $\sigma_A = {\cal C}^2_{A|(BCD)} - \left({\cal C}^2_{A|B} + {\cal C}^2_{A|C} + {\cal C}^2_{A|D} \right) - \left(\left[ \tau_{A|B|C} \right]^{\mu} + \left[ \tau_{A|B|D} \right]^{\mu} + \left[ \tau_{A|C|D} \right]^{\mu} \right)$, and  others are obtained by 
changing the focusing qubit. One can show easily $t_2 (g_3) = 0$. Thus, the measure $t_2$ cannot be excluded as a true four-way entanglement measure.

In Ref. \cite{jin15-1,karmakar16-1} two different negativity-based monogamy relations have been examined. From these relations one can construct the 
following candidates of the four-party entanglement measures:
\begin{equation}
\label{negativity-1}
n_1 = \frac{u_A + u_B + u_C + u_D}{4}
\end{equation}
where $u_A = {\cal N}_{A|(BCD)} - \left( {\cal N}_{A|B} +  {\cal N}_{A|C} +  {\cal N}_{A|D} \right) - \left( \left[ {\cal N}_{A||B|C} \right]^{\nu_1} + \left[ {\cal N}_{A||B|D} \right]^{\nu_1} + \left[ {\cal N}_{A||C|D} \right]^{\nu_1} \right)$ with 
${\cal N}_{I||J|K} \equiv {\cal N}_{I|(JK)} - {\cal N}_{I|J} - {\cal N}_{I|K}$ and
\begin{equation}
\label{negativity-2}
n_2 = \frac{v_A + v_B + v_C + v_D}{4}
\end{equation}
where $v_A = {\cal N}^2_{A|(BCD)} - \left( {\cal N}^2_{A|B} +  {\cal N}^2_{A|C} +  {\cal N}^2_{A|D} \right) - \left( \left[ {\cal N}^2_{A||B|C} \right]^{\nu_2} + \left[ {\cal N}^2_{A||B|D} \right]^{\nu_2} + \left[ {\cal N}^2_{A||C|D} \right]^{\nu_2} \right)$ with
${\cal N}_{I||J|K}^2 \equiv {\cal N}_{I|(JK)}^2 - {\cal N}_{I|J}^2 - {\cal N}_{I|K}^2$. The negativity ${\cal N}$ is defined as 
\cite{vidal02,soojoon03}
\begin{equation}
\label{negativity-3}
{\cal N} (\rho_{AB} ) = || \rho_{AB}^{T_A}|| - 1
\end{equation}
where $||X|| \equiv \mbox{tr} (\sqrt{X X^{\dagger}})$ and the superscript $T_A$ means the partial transposition of $A$-qubit.
Of course other quantities can be obtained by changing the focusing qubit.

The purpose of this paper is to test $t_1$, $t_2$, $n_1$, and $n_2$ by computing them for the rank-$2$ mixture
\begin{equation}
\label{mains}
\rho_4 = p \ket{\mbox{GHZ}_4} \bra{\mbox{GHZ}_4} + (1 - p) \ket{\mbox{W}_4} \bra{\mbox{W}_4}
\end{equation}
where $\ket{\mbox{GHZ}_4}$ is defined in Eq. (\ref{four-maximal}) and 
$\ket{\mbox{W}_4} = (\sigma_x \otimes \sigma_x \otimes \sigma_x \otimes \sigma_x) \ket{\tilde{\mbox{W}}_4}$.
In section II we compute  $t_1$, $t_2$, $n_1$, and $n_2$ for the maximal entangled pure states (\ref{four-maximal}) and $\ket{\mbox{W}_4}$. 
The results are summarized in Table II. It is shown that the negativity-based measures $n_1$ and $n_2$ become negative for $\ket{\Phi_3}$.
In section III we try to compute $t_1$ and $t_2$ for $\rho_4$ by finding the optimal decompositions. For $t_1$ it turns out that  Eq. (\ref{mains}) itself is an 
optimal decomposition. However, we fail to compute $t_2$ because  analytic computation of the residual entanglement is extremely difficult. 
In section IV we compute $n_1$ and $n_2$ for $\rho_4$ in the range $\nu_{1*} \leq \nu_1$ and $\nu_{2*} \leq \nu_2$ by finding the optimal 
decompositions, where 
 \begin{equation}
 \label{nustar}
 \nu_{1*} = \frac{\ln \left( \frac{3 - 3 \sqrt{2} + \sqrt{3}}{6} \right)} {\ln \left( \frac{3}{2} - \sqrt{2} \right)} = 1.02053
 \hspace{.5cm}
 \nu_{2*} = \frac{\ln \left( \frac{\sqrt{2} - 1}{2} \right)} { \ln \left( \sqrt{2} - \frac{5}{4} \right)} = 0.871544.
 \end{equation} 
In this region $n_1 (\mbox{W}_4)$ and $n_2 (\mbox{W}_4)$ become non-negative. In section V a brief conclusion is given. In appendix we try to explain why the 
computation of the residual entanglement is highly difficult.

\section{Computation of $t_1$, $t_2$, $n_1$, and $n_2$ for few special pure states}
In this section we compute $t_1$, $t_2$, $n_1$, and $n_2$ for four-qubit maximally entangled states (\ref{four-maximal}) and W-state $\ket{\mbox{W}_4}$.
The most special case is $\ket{\mbox{GHZ}_4}$, which gives $t_1 = t_2 = n_1 = n_2 = 1$. Since  $\ket{\mbox{W}_4}$ saturates the monogamy 
relations (\ref{tofv}) and (\ref{smonogamy-1}), $t_1$ and $t_2$ of  $\ket{\mbox{W}_4}$ are exactly zero. However,  $\ket{\mbox{W}_4}$ does not 
saturate the negativity-based monogamy relations. It is straightforward\cite{karmakar16-1} to show that $n_1$ and $n_2$ of  $\ket{\mbox{W}_4}$ are 
\begin{eqnarray}
\label{n1n2w4}
&&n_1 (\mbox{W}_4) = \frac{3 + \sqrt{3} - 3 \sqrt{2}}{2} - 3 \left( \frac{3 - 2 \sqrt{2}}{2} \right)^{\nu_1}   \\   \nonumber
&&n_2 (\mbox{W}_4) = \frac{3}{2} (\sqrt{2} - 1) - 3 \left( \frac{4 \sqrt{2} - 5}{4} \right)^{\nu_2}.
\end{eqnarray}
Thus, as we commented, $n_1 (\mbox{W}_4)$ and $n_2 (\mbox{W}_4)$ become non-negative when $\nu_{1*} \leq \nu_1$ and $\nu_{2*} \leq \nu_2$.
 
 For $\ket{\Phi_2}$ it is easy to show that ${\cal C}_{I|(JKL)} = 1$ and ${\cal C}_{I|J} = 0$ for all $\{I, J, K, L\} = \{A, B, C, D\}$. 
 The tripartite states derived from $\ket{\Phi_2} \bra{\Phi_2}$ by tracing over any one-party is given by
 \begin{equation}
 \label{phi2-1}
 \rho_3^{(2)} = \frac{1}{2} \ket{\psi_1} \bra{\psi_1} + \frac{1}{2} \ket{\mbox{W}_3} \bra{\mbox{W}_3}
 \end{equation}
 where $\ket{\psi_1} = (\ket{000} + \sqrt{2} \ket{111}) / \sqrt{3}$ and $\ket{\mbox{W}_3} = (\ket{001} + \ket{010} + \ket{100}) / \sqrt{3}$. 
 In order to compute the residual entanglement of $\rho_3^{(2)}$ let us consider the quantum state  $p \ket{gGHZ} \bra{gGHZ} + (1 - p) \ket{gW} \bra{gW}$,
where $\ket{gGHZ} = a \ket{000} + b \ket{111}$ and $\ket{gW} = c \ket{001} + d \ket{010} + f \ket{100}$.  As the second reference of Ref. \cite{tangle}
has shown, the residual entanglement of this state is exactly  zero when $p \leq p_0 \equiv s^{2/3} / (1 + s^{2/3})$ where $s = 4 c d f/ (a^2 b)$.
Since, for our case, $p_0 = 2/3$ is larger than $1/2$, the residual entanglement of $\rho_3^{(2)}$ is zero. Thus, $t_1$ and $t_2$ of $\ket{\Phi_2}$ are 
\begin{equation}
\label{phi2-2}
t_1 (\Phi_2) = t_2 (\Phi_2) = 1.
\end{equation}
Various negativities of $\ket{\Phi_2}$ can be directly computed and the final expressions are 
\begin{equation}
\label{phi2-3}
{\cal N}_{I|(JKL)} = 1   \hspace{1.0cm}  {\cal N}_{I|(JK)} = \frac{2}{3}  \hspace{1.0cm}  {\cal N}_{I|J} = 0
\end{equation}
for all $\{I, J, K, L\} = \{A, B, C, D\}$. Thus, it is easy to show 
\begin{equation}
\label{phi2-4}
n_1 (\Phi_2) = 1 - 3 \left( \frac{2}{3} \right)^{\nu_1}      \hspace{1.0cm}
n_2 (\Phi_2) = 1 - 3 \left( \frac{4}{9} \right)^{\nu_2}.
\end{equation}
 
 For $\ket{\Phi_3}$ one can show ${\cal C}_{I|(JKL)} = 1$ and ${\cal C}_{I|J} = 0$ for all $\{I, J, K, L\}$. 
 The tripartite states derived from $\ket{\Phi_3} \bra{\Phi_3}$ are 
 \begin{eqnarray}
 \label{phi3-1}
&&\rho_{ACD}^{(3)} = \rho_{BCD}^{(3)} = \frac{1}{2} \ket{\phi_1} \bra{\phi_1} + \frac{1}{2}  \ket{\phi_2} \bra{\phi_2}  \\   \nonumber
&&\rho_{ABC}^{(3)} = \rho_{ABD}^{(3)} = \frac{1}{2} \ket{g_1} \bra{g_1} + \frac{1}{2}  \ket{g_2} \bra{g_2} 
\end{eqnarray}
where
\begin{eqnarray}
\label{phi3-2}
&&\ket{\phi_1} = \frac{1}{\sqrt{2}} (\ket{100} + \ket{111} )  \hspace{1.0cm} \ket{\phi_2} = \frac{1}{\sqrt{2}} (\ket{001} + \ket{010} ) 
                                                                                                                                                                                              \\   \nonumber
&&\ket{g_1} = \frac{1}{\sqrt{2}} (\ket{000} + \ket{111} )  \hspace{1.0cm} \ket{g_2} = \frac{1}{\sqrt{2}} (\ket{001} + \ket{110} ).
\end{eqnarray}  
It is easy to show that the residual entanglements of $\rho_{ACD}^{(3)}$ and $\rho_{BCD}^{(3)}$ are zero because $\ket{\phi_1}$ and 
$\ket{\phi_2}$ are bi-separable. In order to compute the residual entanglement of $\rho_{ABC}^{(3)}$ and $\rho_{ABD}^{(3)}$ let us consider the 
quantum state $p \ket{g_1}\bra{g_1} + (1 - p) \ket{g_2}\bra{g_2}$. As the last reference of Ref. \cite{tangle} has shown, the residual entanglement of 
this state is $(2 p - 1)^2$. Thus, the residual entanglements of  $\rho_{ABC}^{(3)}$ and $\rho_{ABD}^{(3)}$ are also zero, all of which yields
\begin{equation}
\label{phi3-3}
t_1 (\Phi_3) = t_2 (\Phi_3) = 1.
\end{equation}
Various negativities of $\ket{\Phi_3}$ can be computed directly and the final expressions are 
\begin{equation}
\label{phi3-4}
{\cal N}_{I|(JKL)} = 1   \hspace{1.0cm}   {\cal N}_{I|J} = 0
\end{equation}
for all $\{I, J, K, L\} = \{A, B, C, D\}$ and 
\begin{eqnarray}
\label{phi3-5}
&&{\cal N}_{A|(CD)} = {\cal N}_{B|(CD)} = {\cal N}_{C|(AB)} = {\cal N}_{D|(AB)} = 0                         \\    \nonumber
&&{\cal N}_{A|(BC)} = {\cal N}_{A|(BD)} = {\cal N}_{B|(AC)} =  {\cal N}_{B|(AD)} =  {\cal N}_{C|(AD)} =  {\cal N}_{C|(BD)}
=  {\cal N}_{D|(AC)} =  {\cal N}_{D|(BC)} = 1,
\end{eqnarray}   
all of which yields
\begin{equation}
\label{phi3-6}
n_1 (\Phi_3) = n_2 (\Phi_3) = -1.
\end{equation}
All results are summarized in Table II.

\begin{center}
\begin{tabular}{c|cccc} \hline \hline
& $\hspace{.2cm} t_1 \hspace{.2cm}$ &  $\hspace{.2cm} t_2 \hspace{.2cm}$  &  
$\hspace{.2cm} n_1 \hspace{.2cm}$  &  $\hspace{.2cm} n_2 \hspace{.2cm}$   \\  \hline 
$\ket{\mbox{GHZ}_4} \hspace{.2cm}$ & $1$ & $1$ & $1$ & $1$     \\   
$\ket{\Phi_2} \hspace{.2cm}$ & $1$ & $1$ & $1 - 3 \left( \frac{2}{3} \right)^{\nu_1}$ &  $1 - 3 \left( \frac{4}{9} \right)^{\nu_2}$      \\  
$\ket{\Phi_3} \hspace{.2cm}$ & $1$ & $1$ & $-1$   & $-1$              \\  
$\ket{\mbox{W}_4} \hspace{.2cm}$  & $0$  &  $0$  &  $ \frac{3 + \sqrt{3} - 3 \sqrt{2}}{2} - 3 \left( \frac{3 - 2 \sqrt{2}}{2} \right)^{\nu_1}$   &
$\frac{3}{2} (\sqrt{2} - 1) - 3 \left( \frac{4 \sqrt{2} - 5}{4} \right)^{\nu_2}$ \\   \hline  \hline
\end{tabular}

\vspace{0.2cm}
Table II:$t_1$, $t_2$, $n_1$, and $n_2$ of the maximally entangled and $\mbox{W}_4$ states.
\end{center}

\section{Tangle-Based Entanglement Measures for Rank-$2$ Mixture}
In this section we try to compute $t_1$ and $t_2$ for the rank-$2$ mixture $\rho_4$.
For computation of $t_1$ and $t_2$ we have to find an optimal decomposition. 
In order to find the optimal decompositions for  $t_1$ and $t_2$ we define 
\begin{equation}
\label{4qz4}
\ket{Z_4 (p, \varphi)} = \sqrt{p} \ket{\mbox{GHZ}_4} - e^{i \varphi} \sqrt{1-p}  \ket{\mbox{W}_4}.
\end{equation}
Then, one can show that all reduced bipartite states from $\ket{Z_4 (p, \varphi)} \bra{Z_4 (p, \varphi)}$ are equal to 
\begin{eqnarray}
\label{bipartite-1}
\rho_{IJ} = \frac{1}{2} \left(          \begin{array}{cccc}
                                       1  &  -\sqrt{\frac{p (1 - p)}{2}} e^{-i \varphi}  &   -\sqrt{\frac{p (1 - p)}{2}} e^{-i \varphi}   &  0      \\
                                        -\sqrt{\frac{p (1 - p)}{2}} e^{i \varphi} &  \frac{1 - p}{2}  &  \frac{1 - p}{2}  &  0                            \\
                                        -\sqrt{\frac{p (1 - p)}{2}} e^{i \varphi} &  \frac{1 - p}{2}  &  \frac{1 - p}{2}  &  0                            \\ 
                                        0  &  0  &  0  &  p
                                                       \end{array}                                                \right)
                                                       \hspace{.5cm}   (I,J \in \{A, B, C, D\}) 
\end{eqnarray}
in the computational basis. In Eq. (\ref{bipartite-1}) the parties $I$ and $J$ can be chosen any two different parties from 
$\{A, B, C, D \}$.
Although $\rho_{IJ}$ is a rank-$3$ mixture, one can compute its concurrence analytically by following Wootters procedure:
\begin{equation}
\label{bipartite-2}
{\cal C}_{I|J} = \sqrt{\Lambda} - \sqrt{\Lambda_+} - \sqrt{\Lambda_-}
\end{equation}
where 
\begin{eqnarray}
\label{bipartite-3}
&&\Lambda = \frac{1}{12} \left[ (1 + p^2) + 2 (\alpha^2 + \beta^2)^{1/6} \cos \theta \right]    \\   \nonumber
&&\Lambda_{\pm} = \frac{1}{12} \left[ (1 + p^2) - 2 (\alpha^2 + \beta^2)^{1/6} \cos \left( \frac{\pi}{3} \pm \theta\right) \right]
\end{eqnarray}
and 
\begin{eqnarray}
\label{bipartite-4}
&&\alpha = 1 - 9 p + 39 p^2 - 90 p^3 + \frac{231}{2} p^4 -81 p^5 + \frac{47}{2} p^6                          \nonumber     \\
&& \beta = \frac{3 p^2 (1 - p)}{2} \sqrt{3 p (4 - 28 p + 96 p^2 - 147 p^3 + 110 p^4 - 31 p^5)}              \\        \nonumber
&& \theta = \frac{1}{3} \tan^{-1} \left(\frac{\beta}{\alpha} \right).
\end{eqnarray}
\begin{figure}[ht!]
\begin{center}
\includegraphics[height=5.4cm]{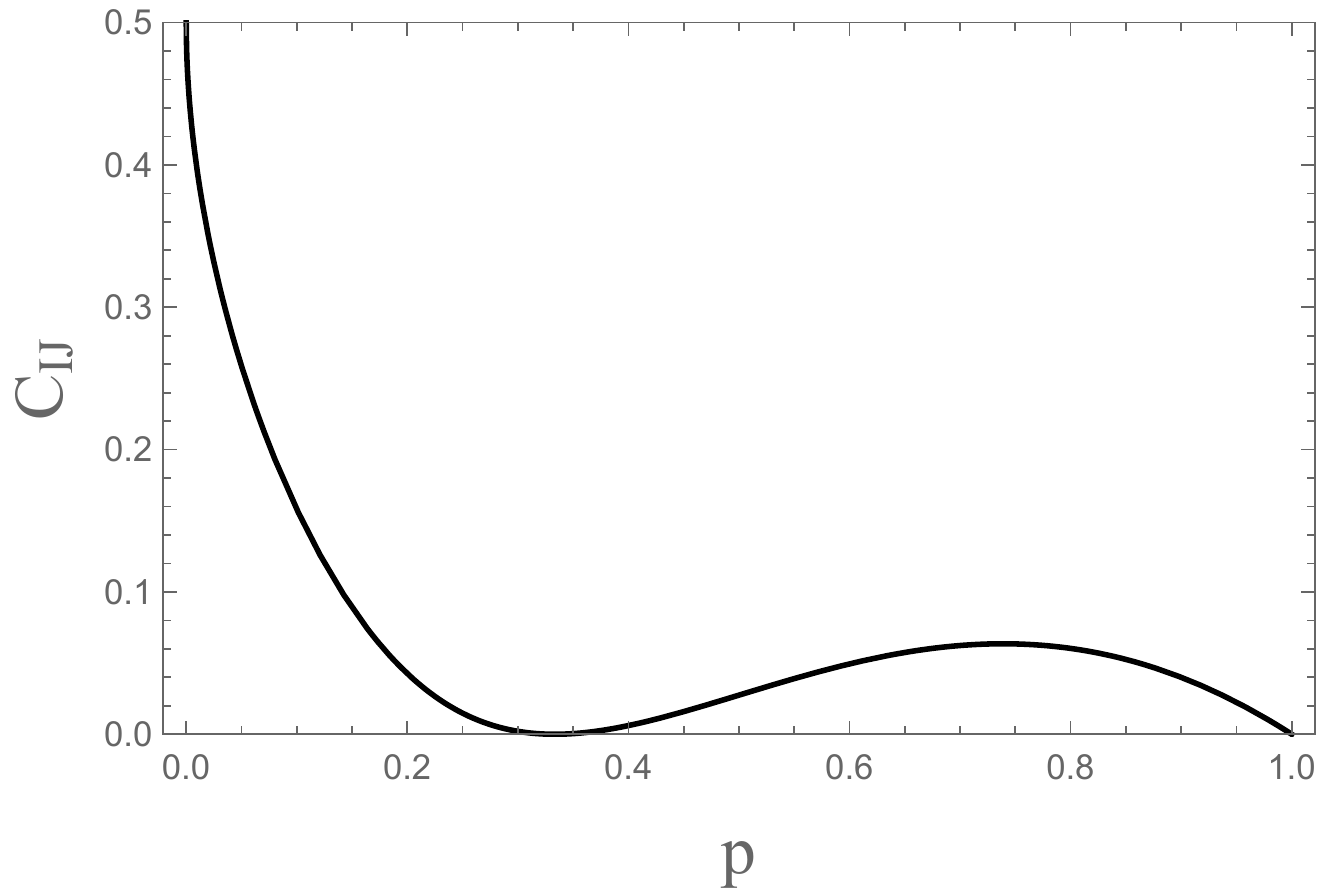}
\includegraphics[height=5.4cm]{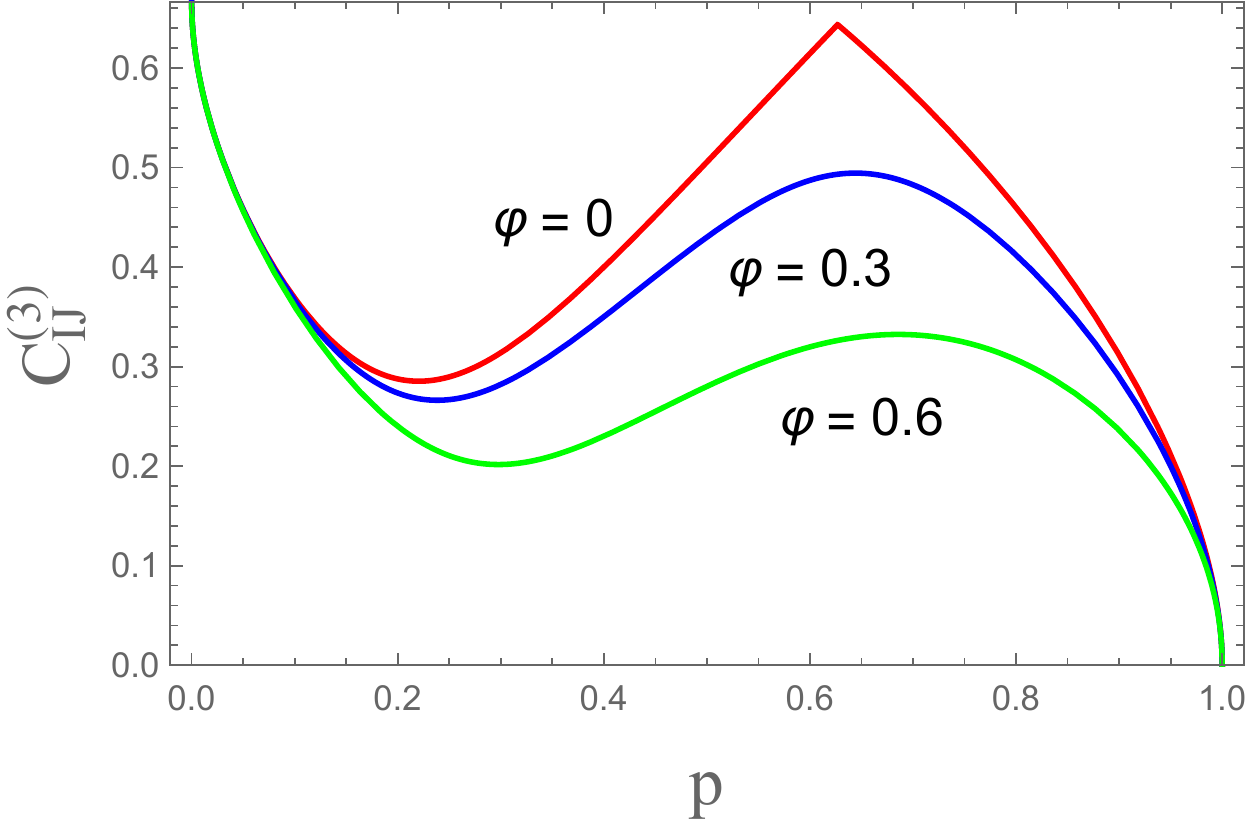}
\caption[fig1]{(Color online) The $p$-dependence of (a) ${\cal C}_{I|J}$ and (b) ${\cal C}_{I|J}^{(3)}$. As Fig. (a) shows, ${\cal C}_{I|J}$ is independent 
of the phase factor $\varphi$ unlike ${\cal C}_{I|J}^{(3)}$. This makes $t_1 (\rho_4)$ trivial. }
\end{center}
\end{figure}
It is interesting to note that ${\cal C}_{I|J} = 0$ at $p = 1 / 3$. Furthermore, it is worthwhile noting that  ${\cal C}_{I|J}$ is independent of the phase factor
$\varphi$. On the contrary, the corresponding concurrence  ${\cal C}_{I|J}^{(3)}$ derived from the three-qubit 
state $\ket{Z_3 (p, \varphi)} = \sqrt{p} \ket{\mbox{GHZ}_3} - e^{i \varphi} \sqrt{1-p}  \ket{\mbox{W}_3}$ is explicitly dependent on $\varphi$.
The $p$-dependence of  ${\cal C}_{I|J}$ and ${\cal C}_{I|J}^{(3)}$ are plotted in Fig. 1. The $\varphi$-dependence of ${\cal C}_{I|J}^{(3)}$ makes the 
three-qubit rank-$2$ mixture $\rho_3 = p \ket{\mbox{GHZ}_3} \bra{\mbox{GHZ}_3} + (1 - p) \ket{\mbox{W}_3} \bra{\mbox{W}_3}$ have the nontrivial 
residual entanglement\cite{tangle}.  As we will show shortly, the $\varphi$-independence of ${\cal C}_{I|J}$ makes $t_1$ of $\rho_4$ to be trivial.

The single qubit states derived from  $\ket{Z_4 (p, \varphi)} \bra{Z_4 (p, \varphi)}$ are all equal to 
\begin{eqnarray}
\label{single-1}
\rho_J = \left(             \begin{array}{cc}
                \frac{3 - p}{4}  &    -\frac{1}{2} \sqrt{\frac{p (1 - p)}{2}} e^{-i \varphi}         \\
                 -\frac{1}{2} \sqrt{\frac{p (1 - p)}{2}} e^{i \varphi}  &  \frac{1 + p }{4}
                                    \end{array}                             \right)
                                    \hspace{.5cm}    (J \in \{A, B, C, D\})
\end{eqnarray}
in the computational basis. Thus, ${\cal C}^2_{I|(JKL)}$ for $\ket{Z_4 (p, \varphi)}$ is 
\begin{equation}
\label{single-2}
{\cal C}^2_{I|(JKL)} = 4 \mbox{det} \rho_I = \frac{3 + p^2}{4}
\end{equation}
for all $I, J, K, L$. The corresponding results derived from $\ket{Z_3 (p, \varphi)}$ is $(8 - 4 p + 5 p^2) / 9$. 
\begin{figure}[ht!]
\begin{center}
\includegraphics[height=5.4cm]{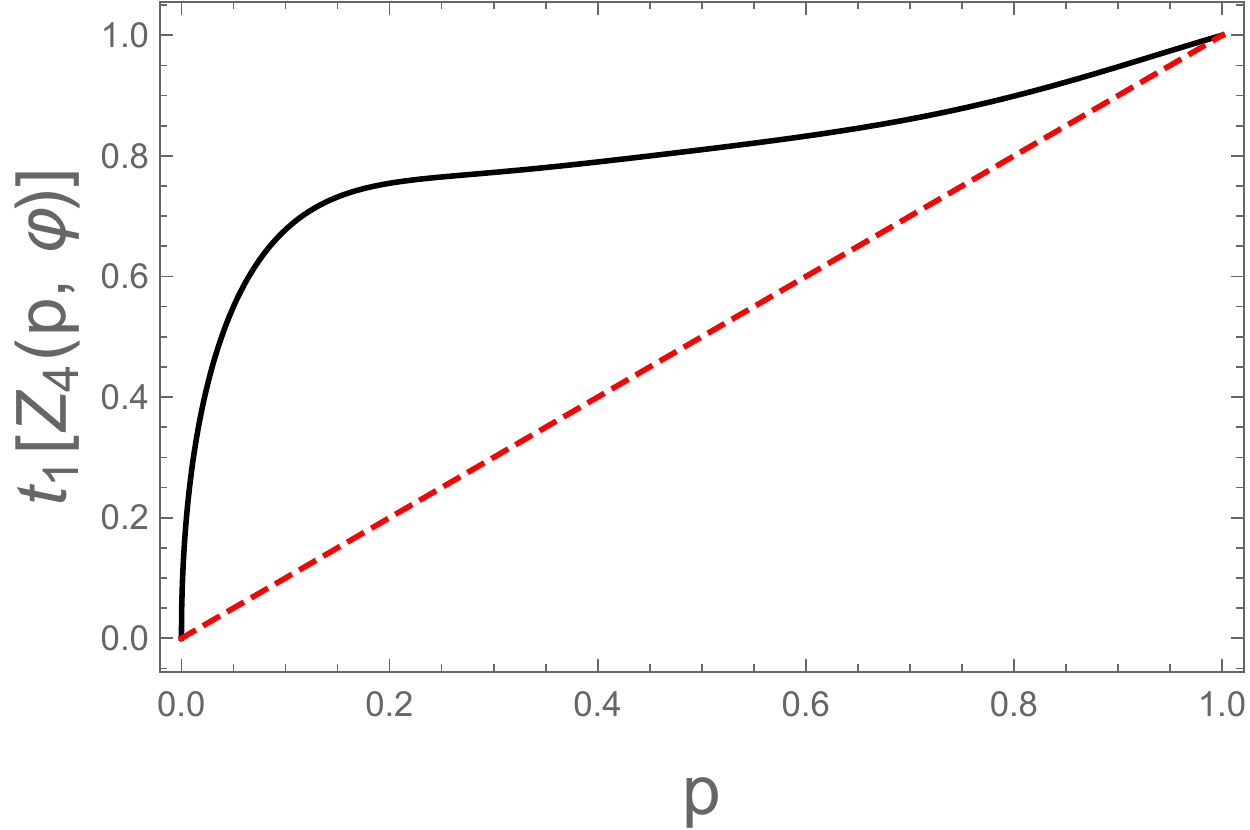}
\includegraphics[height=5.4cm]{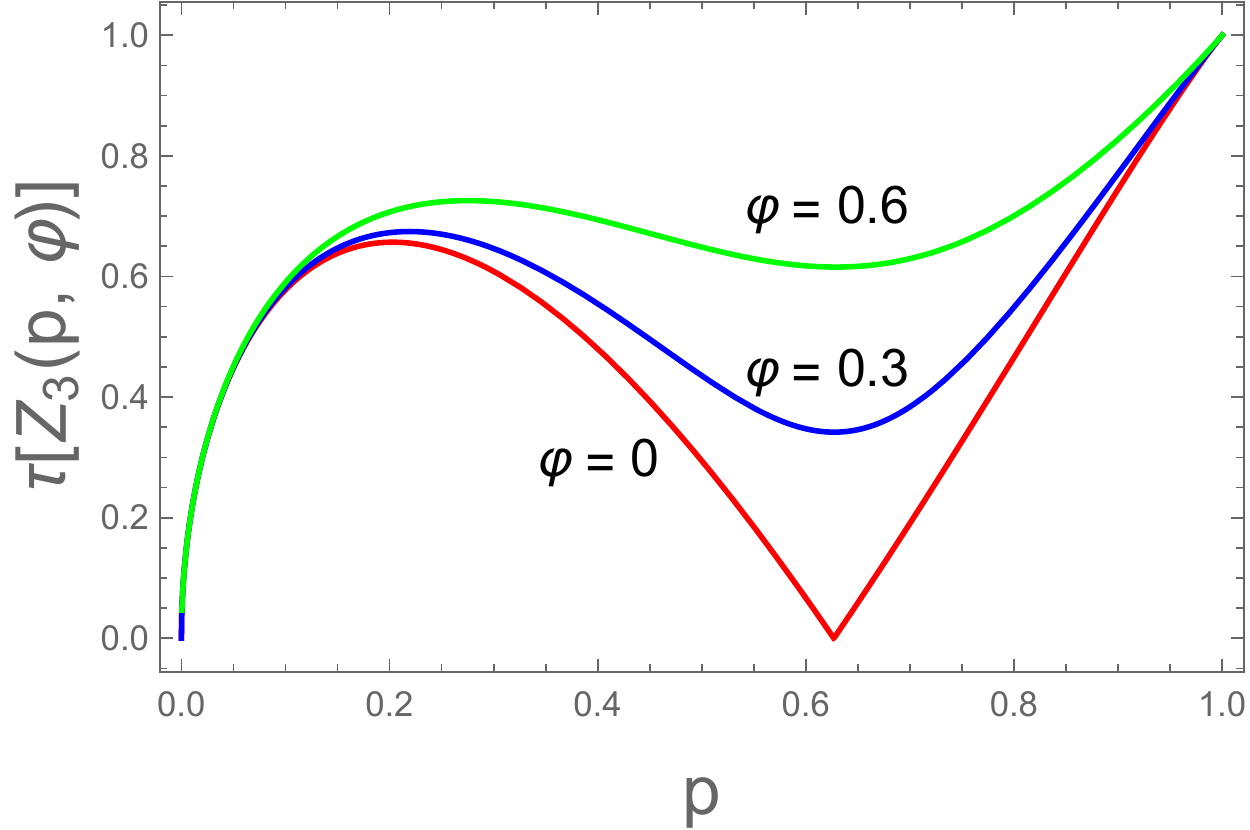}
\caption[fig2]{(Color online) The $p$-dependence of (a) $t_1 \left[Z_4(p, \varphi) \right]$ and (b) $\tau  \left[Z_3(p, \varphi) \right]$. On the contrary to 
$\tau  \left[Z_3(p, \varphi) \right]$  $t_1 \left[Z_4(p, \varphi) \right]$ is independent of the phase factor $\varphi$. }
\end{center}
\end{figure}

Thus, $t_1 \left[Z_4(p, \varphi) \right]$ is 
independent of $\varphi$ as 
\begin{equation}
\label{t1z4}
t_1 \left[Z_4(p, \varphi) \right] = \frac{3 + p^2}{4} - 3 {\cal C}_{I|J}^2.
\end{equation}
The corresponding residual entanglement $\tau  \left[Z_3(p, \varphi) \right]$ for  $\ket{Z_3 (p, \varphi)}$ is dependent on $\varphi$ due to 
${\cal C}_{I|J}^{(3)}$. The $p$-dependence of $t_1 \left[Z_4(p, \varphi) \right]$ and $\tau  \left[Z_3(p, \varphi) \right]$ is plotted in Fig. 2(a)
and Fig. 2(b) respectively. 

Before we calculate $t_1 (\rho_4)$ it seems to be helpful to review briefly how to compute $\tau (\rho_3)$ for 
$\rho_3 = p \ket{\mbox{GHZ}_3} \bra{\mbox{GHZ}_3} + (1 - p) \ket{\mbox{W}_3} \bra{\mbox{W}_3}$. As Fig. 2(b) shows, when $\varphi = 0$ 
$\tau  \left[Z_3(p, \varphi) \right]$ has nontrivial zero at $p = p_0$ with $p_0 \sim 0.627$. Furthermore, $\tau  \left[Z_3(p, \varphi = 0) \right]$
is not convex in the regions $0 \leq p \leq p_0$ and $p_1 \leq p \leq 1$ with  
$p_1 \sim 0.826$. Since $\tau  \left[Z_3(p, \varphi) \right]$ depends on $\varphi$ through only $\cos 3 \varphi$,  
$\tau  \left[Z_3(p_0, \varphi) \right] = 0$ for $\varphi = 0, 2 \pi / 3, 4 \pi / 3$. Thus, at the small concave region it is possible to convexify the 
residual entanglement by making use of $\{\ket{\mbox{W}_3}, \ket{Z_3 (p, \frac{2 \pi}{3} j)} \hspace{.2cm} (j = 0, 1, 2) \}$. At the large concave 
region it is also possible to convexify it  by making use of $\{\ket{\mbox{GHZ}_3}, \ket{Z_3 (p, \frac{2 \pi}{3} j)} \hspace{.2cm} (j = 0, 1, 2) \}$. 

Now, let us return to the four-qubit case. As Fig. 2(a) shows $t_1 \left[Z_4(p, \varphi) \right]$ is not convex at  $0 \leq p \leq p_0$ and $p_1 \leq p \leq 1$
with $p_0 \sim 0.279$ and $p_1 \sim 0.936$. As the three-qubit case it is possible to convexify the entanglement by making use of 
 $\{\ket{\mbox{GHZ}_4}, \ket{Z_4 (p, 0)}, \ket{Z_4 (p, \pi)} \}$ in the large $p$-region. However, it is impossible to convexify it in the small $p$-region
 because  $t_1 \left[Z_4(p_0, 0) \right] \neq 0$. The only way to obtain the convex result in the entire range of $p$ is 
 \begin{equation}
 \label{t1-final}
 t_1 (\rho_4) = p.
 \end{equation}
 As Fig. 2(a) shows obviously as a dashed line this is a convex hull of $t_1 \left[Z_4(p, \varphi) \right]$. Thus the optimal decomposition for $t_1$ is nothing
 but the spectral decomposition (\ref{mains}) itself.  
 
 In order to compute $t_2 (\rho_4)$ we should compute the residual entanglement for the three-qubit states reduced from $\ket{Z_4(p, \varphi)}$. One 
 can show that all tripartite states derived by tracing over single qubit are equal to 
 \begin{equation}
 \label{three-1}
 \rho_{IJK} = \lambda \ket{\psi_+} \bra{\psi_+} + (1 - \lambda) \ket{\psi_-} \bra{\psi_-}  
 \hspace{.5cm} (I,J,K \in \{A, B, C, D\})
 \end{equation}
 where
 \begin{eqnarray}
 \label{three-2}
 && \lambda = \frac{2 + \sqrt{1 - p^2}}{4}                            \\    \nonumber
 &&\ket{\psi_{\pm}} = \frac{1}{N_{\pm}} \left[ \mu_{\pm} \ket{000} - e^{-i \varphi} \ket{111}
        - \nu_{\pm} e^{i \varphi} \left( \ket{001} + \ket{010} + \ket{100} \right)  \right]
\end{eqnarray}
with
\begin{eqnarray}
\label{three-3}
&&N_{\pm}^2 = \frac{(1 + p) (3 - p) \pm (3 + p) \sqrt{1 - p^2}}{2 p^2}       \hspace{.5cm}
\mu_{\pm} = \frac{2 (1 - p) \pm \sqrt{1 - p^2}}{\sqrt{2 p (1 - p)}}                                         \\    \nonumber
&& \hspace{3.0cm} \nu_{\pm} = \frac{(3 + p) (1 - p) \pm (3 - p) \sqrt{1 - p^2}}{2 p (2 \pm \sqrt{1 - p^2})}.
\end{eqnarray}
The residual entanglement for $\ket{\psi_{\pm}}$ is 
\begin{equation}
\label{three-4}
\tau (\psi_{\pm}) = \frac{4}{N_{\pm}^4} \sqrt{\mu_{\pm}^4 + 16 \nu_{\pm}^6 + 8 \mu_{\pm}^2 \nu_{\pm}^3 \cos 4 \varphi}.
\end{equation}
Thus, the spectral decomposition (\ref{three-1}) indicates that the residual entanglement for $\rho_{IJK}$ satisfies
\begin{equation}
\label{three-5}
\tau (\rho_{IJK}) \leq \lambda \tau (\psi_+) + (1 - \lambda) \tau (\psi_-).
\end{equation}
However, the analytic computation of the residual entanglement for $\rho_{IJK}$ is highly difficult  even though it is rank-$2$ tensor.
In appendix we try to describe why it is highly difficult. Therefore, we fail to compute $t_2 (\rho_4)$ analytically.

\section{Negativity-Based Entanglement Measures for Rank-$2$ Mixture}
In this section we try to compute $n_1$ and $n_2$ for $\rho_4$. We consider only the regions $\nu_{1*} \leq \nu_1 \leq \infty$ and 
 $\nu_{2*} \leq \nu_2 \leq \infty$, where $\nu_{1*}$ and $\nu_{2*}$ are given in Eq. (\ref{nustar}).
When $\nu_1 = \nu_{1*}$ and $\nu_2 = \nu_{2*}$, $n_1(W_4) = n_2(W_4) = 0$ exactly. When $\nu_1 = \nu_2 = \infty$, $n_1$ and $n_2$ for 
 $\ket{\mbox{W}_4}$ become
 \begin{equation}
 \label{n1n2w4}
n_1 (\mbox{W}_4) = \frac{3 + \sqrt{3} - 3 \sqrt{2}}{2} = 0.244705    \hspace{.5cm}
n_2 (\mbox{W}_4) = \frac{3}{2} (\sqrt{2} - 1) = 0.62132.
\end{equation}
Of course, $n_1$ and $n_2$ for $\ket{\mbox{GHZ}_4}$ are unity regardless of $\nu_1$ and $\nu_2$. 
  
In order to find the optimal decompositions for $n_1 (\rho_4)$ and $n_2 (\rho_4)$ we re-consider $\ket{Z_4 (p, \varphi)}$ defined in Eq. (\ref{4qz4}).
By direct calculation one can show straightforwardly
\begin{equation}
\label{totalbipartite}
{\cal N}_{I|(JKL)} = \frac{1}{2} \sqrt{3 + p^2}
\end{equation}
where $I, J, K, L$ are any one of $\{A, B, C, D\}$. 

Using Eq. (\ref{bipartite-1}) one can also show that any bipartite negativity ${\cal N}_{I|J}$ for 
$\ket{Z_4 (p, \varphi)}$ is
\begin{equation}
\label{nbipartite-1}
{\cal N}_{I|J} = \sqrt{\lambda} + \sqrt{\lambda_+} + \sqrt{\lambda_-} - \frac{3 + p}{4}
\end{equation}
where
\begin{equation}
\label{nbipartite-2}
\lambda = \frac{1}{48} \left[ (7 + 2 p - p^2) + 4 r_0 \cos \theta_0 \right]    \hspace{.5cm}
\lambda_{\pm} = \frac{1}{48} \left[ (7 + 2 p - p^2) - 4 r_0 \cos \left( \frac{\pi}{3} \pm \theta_0 \right) \right]
\end{equation}
with 
\begin{eqnarray}
\label{nbipartite-3}
&&r_0 = (\alpha_0^2 + \beta_0^2)^{1/6}   \hspace{.5cm} \theta_0 = \frac{1}{3} \tan^{-1} \left( \frac{\beta_0}{\alpha_0} \right)   \nonumber   \\
&&\alpha_0 = 17 + 147 p - 153 p^2 - 428 p^3 + 729 p^4 - 447 p^5 + 127 p^6                                            \\      \nonumber
&&\beta_0 = 3 \sqrt{3} (1 - p + 5 p^2 - 7 p^3 + 2 p^4) \sqrt{2 + 4 p - 71 p^2 + 214 p^3 - 129 p^4}.
\end{eqnarray}
Like the concurrence discussed in the previous section ${\cal N}_{I|J}$ becomes zero when $p = 1/3$ and $p = 1$. 

\begin{figure}[ht!]
\begin{center}
\includegraphics[height=8cm]{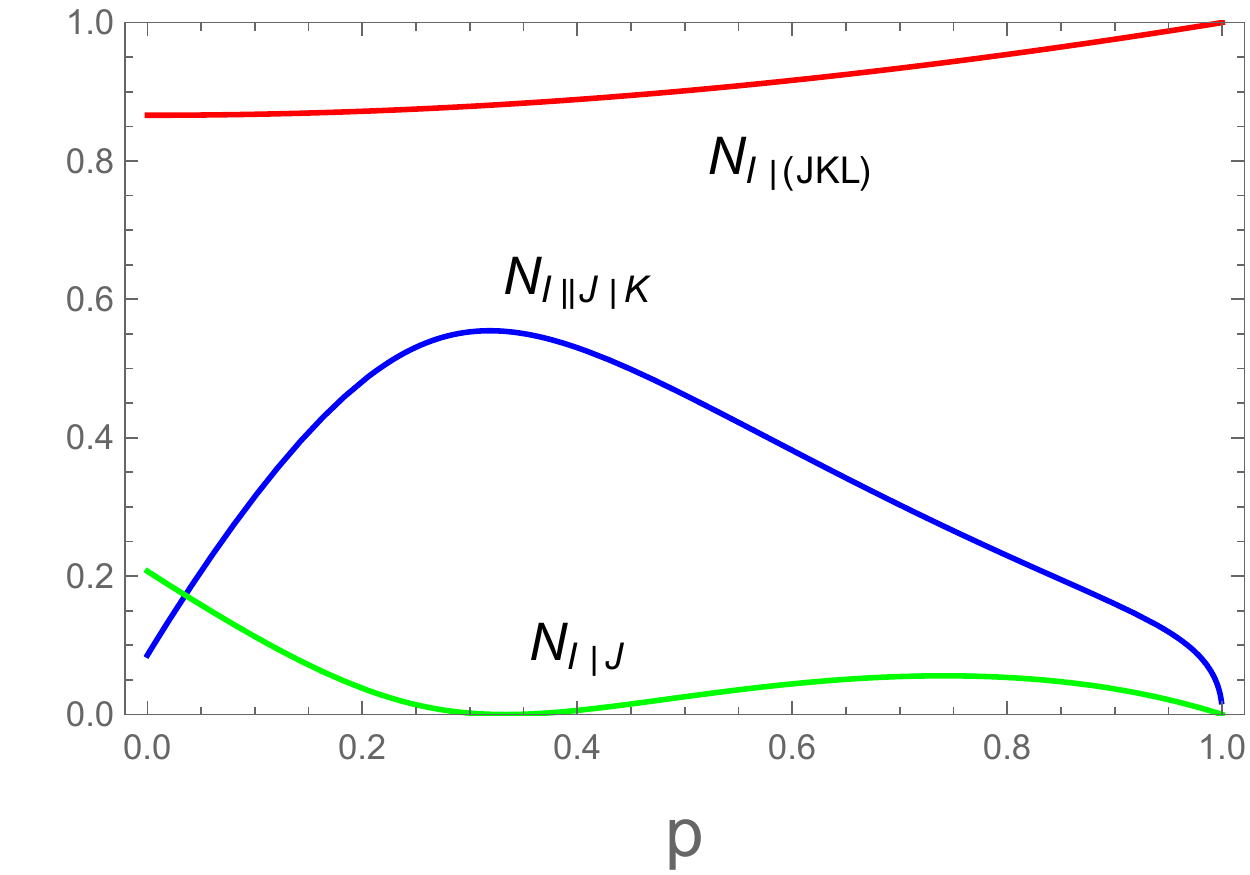}
\caption[fig3]{(Color online) The $p$-dependence of ${\cal N}_{I|(JKL)}$, ${\cal N}_{I||J|K}$, and ${\cal N}_{I|J}$ for $\ket{Z(p, \varphi}$. It is shown that 
all negativities are independent of the phase factor $\varphi$.}
\end{center}
\end{figure}

Finally, we compute ${\cal N}_{I||(JK)}$ for all $I, J, K \in \{A, B, C, D\}$. Using Eq. (\ref{three-1}) one can compute the non-zero eigenvalues of 
$\left(\rho_{IJK}^{T_I} \right) \left(\rho_{IJK}^{T_I} \right)^{\dagger}$. One of them is $p^2 / 4$ and the remaining five non-zero eigenvalues 
can be obtained by solving the quintic equation. Thus, it is possible to compute ${\cal N}_{I||(JK)}$  numerically. After obtaining ${\cal N}_{I||(JK)}$, 
one can compute ${\cal N}_{I||J|K}$ and  ${\cal N}_{I||J|K}^2$ by making use of 
${\cal N}_{I||J|K} = {\cal N}_{I||(JK)} - ({\cal N}_{I|J} + {\cal N}_{I|K})$ and 
${\cal N}_{I||J|K}^2 = {\cal N}_{I||(JK)}^2 - ({\cal N}_{I|J}^2 + {\cal N}_{I|K}^2)$. It is worthwhile noting that all negativities are independent of 
the phase angle $\varphi$. Thus, $n_1 (\rho_4)$ and $n_2 (\rho_4)$ are independent of $\varphi$. 
In Fig. 3 we plot the $p$-dependence of ${\cal N}_{I|(JKL)}$, 
${\cal N}_{I||J|K}$, and ${\cal N}_{I|J}$. 

In Fig. 4(a) we plot the $p$-dependence of $n_1 [Z(p, \varphi)]$ for $\ket{Z_4(p, \varphi)}$ when $\nu_1 = \nu_{1*}$ (red dashed line) and 
 $\nu_1 = \infty$ (blue dotted line). When $\nu_1 = \nu_{1*}$, $n_1 [Z(p, \varphi)]$ becomes negative at $0 < p < p_0$, where $p_0 = 0.749596$.
 Since, however, $n_1 (\mbox{W}_4) = 0$ at $\nu_1 = \nu_{1*}$, one can choose the optimal decomposition for $\rho_4 (p)$ in this region as 
 \begin{equation}
 \label{optimaln-1}
 \rho_4(p) = \frac{p}{2 p_0} \left[ \ket{Z_4 (p_0, 0)} \bra{Z_4 (p_0, 0)} +  \ket{Z_4 (p_0, \pi)} \bra{Z_4 (p_0, \pi)} \right] 
 + \left(1 - \frac{p}{p_0}\right) \ket{\mbox{W}_4} \bra{\mbox{W}_4},
 \end{equation}
which gives $n_1 (\rho_4) = 0$ at $0 \leq p \leq p_0$. At $p_0 \leq p \leq 1$ the optimal decomposition for $\rho_4 (p)$ is 
\begin{equation}
\label{optimaln-2}
\rho_4 (p) = \frac{1}{2}  \left[ \ket{Z_4 (p, 0)} \bra{Z_4 (p, 0)} +  \ket{Z_4 (p, \pi)} \bra{Z_4 (p, \pi)} \right],
\end{equation}
 which gives $n_1 (\rho_4) = n_1 [Z(p, \varphi)]$ at $p_0 \leq p \leq 1$. Since $ n_1 [Z(p, \varphi)]$ is convex in this region, we do not need to 
 convexify it. Thus, our result for $n_1 (\rho_4)$ at $\nu_1 = \nu_{1*}$ can be expressed as
 \begin{eqnarray}
 \label{n1nustar-final}
 n_1 (\rho_4) = \left\{        \begin{array}{cc}
                                    0    &    \hspace{1.0cm}0 \leq p \leq p_0 = 0.749596    \\
                                    n_1[Z(p, 0)]  &  \hspace{1.0cm}p_0 \leq p \leq 1.
                                          \end{array}             \right.
\end{eqnarray}
This is plotted in Fig. 4(a) as a red (lower) solid line.
 
  When $\nu_1 = \infty$, $n_1[Z(p, \varphi)]$ is not convex at the region $0 \leq p \leq p_*$ with $p_* \approx 0.475$. Thus, we have to convexify it 
 at the region $0 \leq p \leq p_1$ with $p_1 > p_*$. We will fix $p_1$ later. At the region $0 \leq p \leq p_1$ we choose an optimal decomposition for 
 $\rho_4(p)$ as 
 \begin{equation}
 \label{optimaln-3}
 \rho_4(p) = \frac{p_1 - p}{p_1} \ket{\mbox{W}_4} \bra{\mbox{W}_4} + \frac{p}{2 p_1} 
\left[ \ket{Z_4 (p_1, 0)} \bra{Z_4 (p_1, 0)} +  \ket{Z_4 (p_1, \pi)} \bra{Z_4 (p_1, \pi)} \right].
\end{equation}
From Eq. (\ref{optimaln-3}) $n_1(\rho_4)$ becomes $g(p)$, where
\begin{equation}
\label{optimaln-4}
g(p) = \frac{3 + \sqrt{3} - 3 \sqrt{2}}{2} \frac{p_1 - p}{p_1} + \frac{p}{p_1} n_1 [Z(p_1, 0)].
\end{equation}
Then, $p_1$ is determined by $\partial g(p, p_1) / \partial p_1 = 0$, which gives $p_1 \approx 0.84$. 
Thus, finally $n_1 (\rho_4)$ at $\nu_1 = \infty$ is given by 
\begin{eqnarray}
 \label{n1inf-final}
 n_1 (\rho_4) = \left\{        \begin{array}{cc}
                                    g(p)    &    \hspace{1.0cm}0 \leq p \leq p_1 \approx 0.84     \\
                                    n_1[Z(p, 0)]  &  \hspace{1.0cm}p_1 \leq p \leq 1.
                                          \end{array}             \right.
\end{eqnarray}
This is plotted in Fig. 4(a) as a blue (upper) solid line.

\begin{figure}[ht!]
\begin{center}
\includegraphics[height=5.4cm]{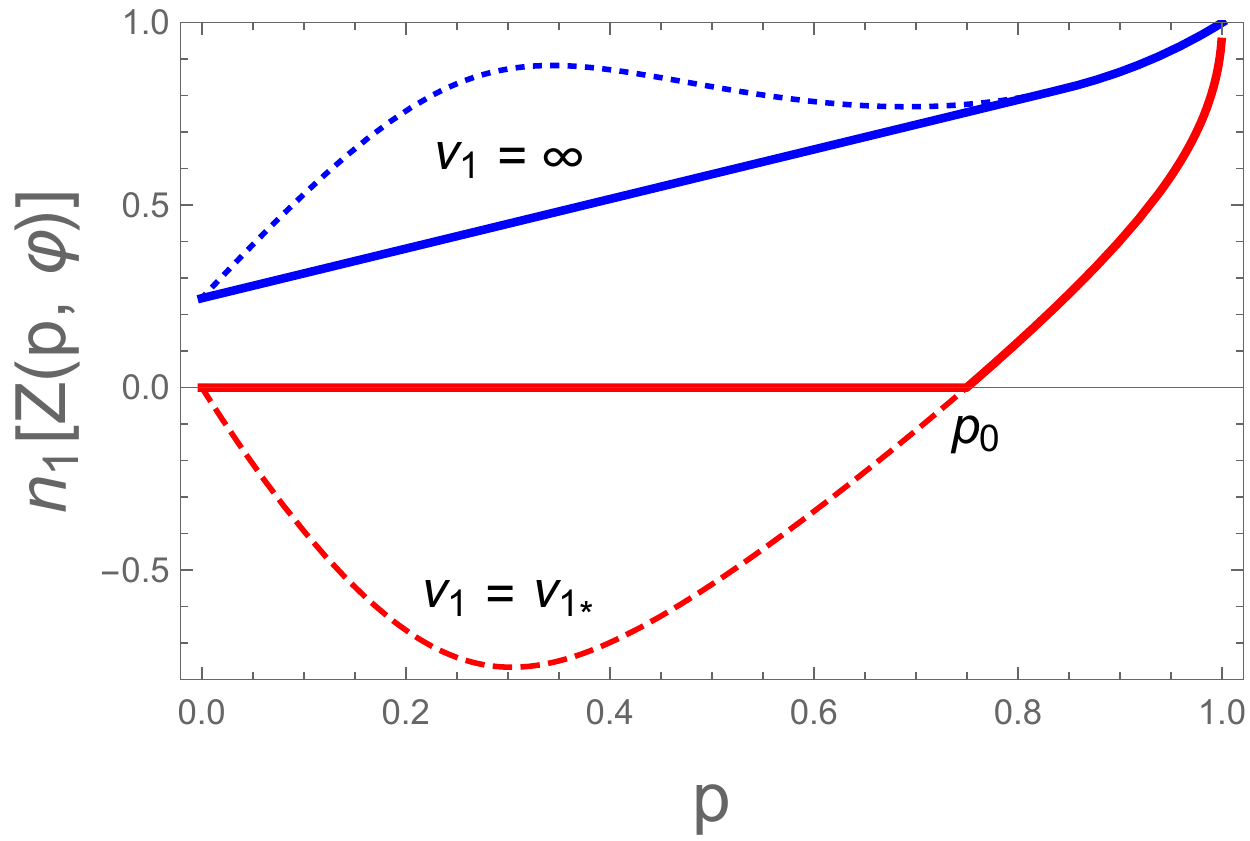}
\includegraphics[height=5.4cm]{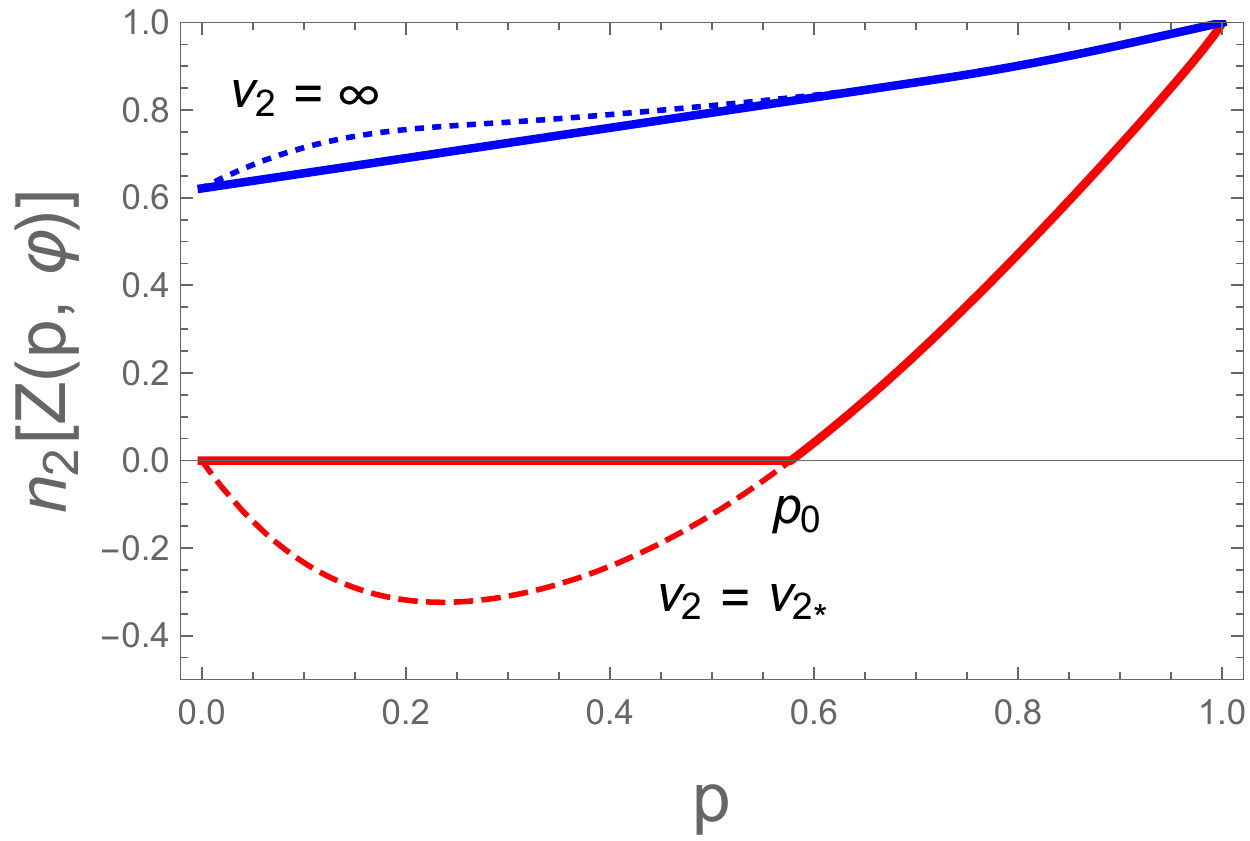}
\caption[fig4]{(Color online) The $p$-dependence of (a)  $n_1 [Z(p, \varphi)]$ (dashed and dotted) and $n_1 (\rho_4)$ (solid) at $\nu_1 = \nu_{1*}$ (red (lower) solid) and 
$\nu_1 = \infty$ (blue (upper) solid) (b)  $n_2 [Z(p, \varphi)]$ (dashed and dotted ) and $n_2 (\rho_4)$ (solid) at $\nu_2 = \nu_{2*}$ (red (lower) solid) and 
$\nu_2 = \infty$ (blue (upper) solid). }
\end{center}
\end{figure}

In Fig. 4(b) we plot the $p$-dependence of $n_2 [Z(p, \varphi)]$ for $\ket{Z_4 (p, \varphi)}$ when $\nu_2 = \nu_{2*}$ (red dashed line) and 
$\nu_2 = \infty$ (blue dotted line). When  $\nu_2 = \nu_{2*}$, $n_2 [Z(p, \varphi)]$ becomes negative at the region $0 \leq p \leq p_0$, where
$p_0 = 0.57731$. Following the similar procedure in the case of $\nu_1 = \nu_{1*}$ one can derive $n_2 (\rho_4)$ as 
\begin{eqnarray}
 \label{n2nustar-final}
 n_2 (\rho_4) = \left\{        \begin{array}{cc}
                                    0    &    \hspace{1.0cm}0 \leq p \leq p_0 = 0.57731    \\
                                    n_2[Z(p, 0)]  &  \hspace{1.0cm}p_0 \leq p \leq 1.
                                          \end{array}             \right.
\end{eqnarray}
This is plotted in Fig. 4(b) as a red (lower) solid line. 

For $\nu_2 = \infty$ case $n_2 [Z(p, \varphi)]$ is not convex at $0 \leq p \leq p_{1*}$ and $p_{2*} \leq p \leq 1$, where $p_{1*} \approx 0.25$ and 
$p_{2*} \approx 0.95$. Thus, we have to convexify $n_2 (\rho_4)$ in the small-$p$ and large-$p$ regions. First we choose a small-$p$ region
$0 \leq p \leq p_1$ with $p_{1*} \leq p_1 \leq p_{2*}$. The parameter $p_1$ will be fixed later. In this region we choose the optimal decomposition as
Eq. (\ref{optimaln-3}). Then, $n_2 (\rho_4)$ becomes $f_I (p)$, where 
\begin{equation}
\label{optimaln-5}
f_I (p) = \frac{3}{2} (\sqrt{2} - 1) \frac{p_1 - p}{p_1} + \frac{p}{p_1} n_2 [Z(p_1, 0)].
\end{equation}
Then, $p_1$ is fixed by $\partial f_I(p, p_1) / \partial p_1 = 0$, which gives $p_1 \approx 0.72$. Next, we consider the large-$p$ region
$p_2 \leq p \leq 1$ with $p_1 \leq p_2 \leq p_{2*}$. In this region the optimal decomposition can be chosen as 
\begin{equation}
\label{optimaln-6}
\rho_4 (p) = \frac{p - p_2}{1 - p_2} \ket{\mbox{GHZ}_4} \bra{\mbox{GHZ}_4} + 
\frac{1 - p}{2 (1 - p_2)} \left[ \ket{Z_4 (p_2, 0)} \bra{Z_4 (p_2, 0)} +  \ket{Z_4 (p_2, \pi)} \bra{Z_4 (p_2, \pi)} \right].
\end{equation}
Thus, $n_2 (\rho_4)$ becomes $f_{II} (p)$ in this region, where 
\begin{equation}
\label{optimaln-7}
f_{II} (p) = \frac{p - p_2}{1 - p_2} + \frac{1 - p}{1 - p_2} n_2 [Z(p_2, 0)].
\end{equation}
The parameter  $p_2$ is fixed by $\partial f_{II}(p, p_2) / \partial p_2 = 0$, which gives $p_2 \approx 0.92$. Thus, the final expression 
$n_2 (\rho_4)$ for $\nu_2 = \infty$ case can be written in a form
\begin{eqnarray}
\label{n2inf-final}
 n_2 (\rho_4) = \left\{        \begin{array}{cc}
                                    f_I (p)    &    \hspace{1.0cm}0 \leq p \leq p_1 \approx 0.72    \\
                                    n_2[Z(p, 0)]  &  \hspace{1.0cm}p_1 \leq p \leq p_2 \approx 0.92   \\
                                    f_{II} (p)  &    \hspace{1.0cm} p_2 \leq p \leq 1.
                                          \end{array}             \right.
\end{eqnarray}
This is plotted as a blue (upper) solid line in Fig. 4(b).

\section{Conclusions}
In this paper we compute the monogamy-motivated four-party measures $n_1$, $n_2$, $t_1$, and $t_2$ for the rank-$2$ mixtures $\rho_4$ given in Eq. (\ref{mains}).
It turns out that $t_1 (\rho_4)$ is trivial and the corresponding optimal decomposition is equal to the spectral decomposition. Probably, this triviality 
is a sign of the fact that monogamy relation (\ref{tofv}) is not sufficiently tight, which means that $t_1$ is not a true four-way entanglement measure.
We fail to compute $t_2 (\rho_4)$ analytically because it is highly difficult to compute the residual entanglement for the rank-$2$ state (\ref{three-1}), 
which is a tripartite state reduced from $\ket{Z_4(p, \varphi)} \bra{Z_4(p, \varphi)}$. This difficulty is discussed in the appendix.

We also compute $n_1$ and $n_2$ for $\rho_4$ when $\nu_j = \nu_{j*} \hspace{.2cm} (j = 1, 2)$ or $\infty$. When $\nu_1 = \nu_{1*}$ the final expression of 
$n_1 (\rho_4)$ is Eq. (\ref{n1nustar-final}) and the corresponding optimal decompositions are (\ref{optimaln-1}) in $0 \leq p \leq p_0$ and (\ref{optimaln-2}) in 
$p_0 \leq p \leq 1$, where $p_0 \sim 0.749596$. When $\nu_1 = \infty$, the final expression of $n_1 (\rho_4)$ is Eq. (\ref{n1inf-final}) and the 
corresponding optimal decompositions are (\ref{optimaln-3}) in $0 \leq p \leq p_1$ and (\ref{optimaln-2}) in $p_1 \leq p \leq 1$, where 
$p_1 \sim 0.84$. When $\nu_2 = \nu_{2*}$ and $\nu_2 = \infty$, the final expressions of $n_2 (\rho_4)$ are Eq. (\ref{n2nustar-final}) and 
Eq. (\ref{n2inf-final}), respectively and the corresponding optimal decompositions can be found in the previous section. As Table II shows $n_1$ and $n_2$ are 
not always non-negative for all four-qubit pure states. This means that the corresponding negativity-based monogamy relations discussed in 
Ref.\cite{jin15-1,karmakar16-1}  do not always hold regardless of the power factor $\nu_1$ and $\nu_2$. 

It is most important for us to check whether or not $t_2$ is a true four-way entanglement measure when the power factor $\mu_3$ is chosen appropriately. 
As we mentioned we fail to check this fact in this paper due to the difficulty for the analytic computation of the residual entanglement of the 
tripartite reduced state (\ref{three-1}). We hope to discuss this issue again in the near future.

{\bf Acknowledgement}:
On April 16, 2014 the ferry Sewol has sunk into the South Sea of Korea. Due to this disaster 304 people died and, 9 of them are still missing. We would like to dedicate this paper to all victims of this accident.

\newpage

\begin{appendix}{\centerline{\bf Appendix A}}

\setcounter{equation}{0}
\renewcommand{\theequation}{A.\arabic{equation}}

In this appendix we try to explain why the analytic computation of the residual entanglement for $\rho_{IJK}$ in Eq. (\ref{three-1}) is difficult by introducing a 
simpler rank-$2$ quantum state
\begin{equation}
\label{appen-1}
\Pi = p \ket{\psi_1} \bra{\psi_1} + (1 - p) \ket{\psi_2} \bra{\psi_2}
\end{equation}
where
\begin{equation}
\label{appen-2}
\ket{\psi_1} = \frac{1}{\sqrt{2}} \left( \ket{\mbox{GHZ}_3} + \ket{\mbox{W}_3} \right)      \hspace{1.0cm}
\ket{\psi_2} = \frac{1}{\sqrt{2}} \left( \ket{\mbox{GHZ}_3} - \ket{\mbox{W}_3} \right).
\end{equation}
In spite of its simpleness $\Pi$ has a same structure with $\rho_{IJK}$. The residual entanglement of $\ket{\psi_1}$ and $\ket{\psi_2}$ are
\begin{equation}
\label{appen-3}
\tau_3 (\psi_1)  = \frac{8 \sqrt{6} + 9}{36} = 0.794331        \hspace{1.0cm}
\tau_3 (\psi_2)  = \frac{8 \sqrt{6} - 9}{36} = 0.294331.
\end{equation}
Thus, $\tau_3 (\Pi)$ has an upper bound as 
\begin{equation}
\label{appen-4}
\tau_3 (\Pi) \leq \tau_3^{\max} = \frac{p}{2} +  \frac{8 \sqrt{6} - 9}{36}.
\end{equation}

In order to compute $\tau_3 (\Pi)$ we define the superposed state
\begin{equation}
\label{appen-5}
\ket{Z(p, \varphi)} = \sqrt{p} \ket{\psi_1} - e^{i \varphi} \sqrt{1 - p} \ket{\psi_2}.
\end{equation}
The residual entanglement $\tau_3 (p, \varphi)$ of $\ket{Z(p, \varphi)}$ can be written as 
\begin{equation}
\label{appen-6}
\tau_3 (p, \varphi) = 4 p^2 \bigg| \frac{1 - z}{2} \bigg| \bigg| \frac{1}{8} (1 - z)^3 + \frac{2}{3 \sqrt{6}} (1 + z)^3 \bigg|
\end{equation}
where
\begin{equation}
\label{appen-7}
z = e^{i \varphi} \sqrt{\frac{1 - p}{p}}.
\end{equation}
Another useful expression of $\tau_3 (p, \varphi)$ is 
\begin{eqnarray}
\label{appen-8}
&&\tau_3 (p, \varphi)      \\     \nonumber
&&= 2 \sqrt{\left( 1 - 2 \sqrt{p (1 - p)} \cos \varphi \right) \left( f_0 (p) + f_1 (p) \cos \varphi + f_2(p) \cos 2 \varphi + 
f_3 (p) \cos 3 \varphi \right)}
\end{eqnarray}
where
\begin{eqnarray}
\label{appen-9}
&&f_0 (p) = \frac{155}{1728} (1 + 6 p - 6 p^2) + \frac{2 p -1}{6 \sqrt{6}} (1 - 10 p + 10 p^2)               \nonumber     \\
&&f_1 (p) = \frac{101}{288} \sqrt{p (1 - p)} (1 + p - p^2)                           \\              \nonumber
&&f_2 (p) = 6 p (1 - p) \left( \frac{155}{1728} + \frac{2 p - 1}{6 \sqrt{6}}  \right)                \\    \nonumber
&&f_3 (p) = \frac{101}{864} \sqrt{p^3 (1 - p)^3}.
\end{eqnarray}

\begin{center}
\begin{tabular}{c|ccc} \hline \hline
$\hspace{.5cm} \varphi \hspace{.5cm}$ &  $\hspace{.5cm} \pi \hspace{.5cm}$  &  
$\hspace{.5cm} 0 \hspace{.5cm}$  &  $\hspace{.5cm} \pi \pm \varphi_0 \hspace{.5cm}$   \\  \hline 
$\hspace{.5cm}p\hspace{.5cm}$ & $\hspace{.5cm}p_1 = 0.0163588\hspace{.5cm}$ & $\hspace{.5cm}p_2 = 0.5\hspace{.5cm}$ & $\hspace{.5cm}p_3 = 0.74182\hspace{.5cm}$      \\ \hline \hline  
\end{tabular}

\vspace{0.2cm}
Table III:Nontrivial zeros of $\tau_3(p, \varphi)$  with $\varphi_0 = 1.27672$.
\end{center}

From Eq. (\ref{appen-6}) one can show that $\tau_3 (p, \varphi)$ becomes zero at particular $p$ and $\varphi$. These nontrivial zeros are 
summarized in Table III. From Eq. (\ref{appen-8}) one can show that $\tau_3 (p, \varphi)$ has a symmetry
\begin{equation}
\label{appen-10}
\tau_3 (p, n \pi + \varphi_0) = \tau_3 (p, n \pi - \varphi_0)
\end{equation}
for all integer $n$.

\begin{figure}[ht!]
\begin{center}
\includegraphics[height=5.4cm]{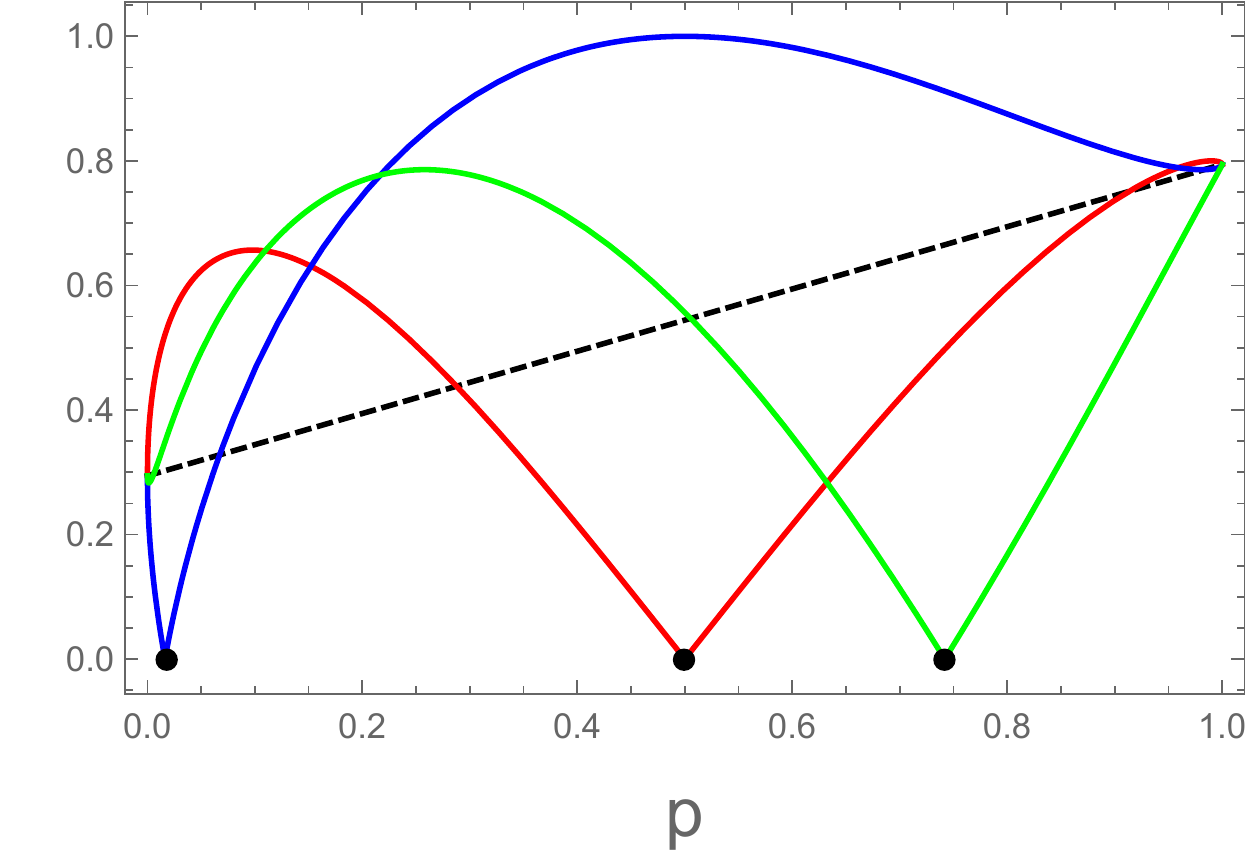}
\includegraphics[height=5.4cm]{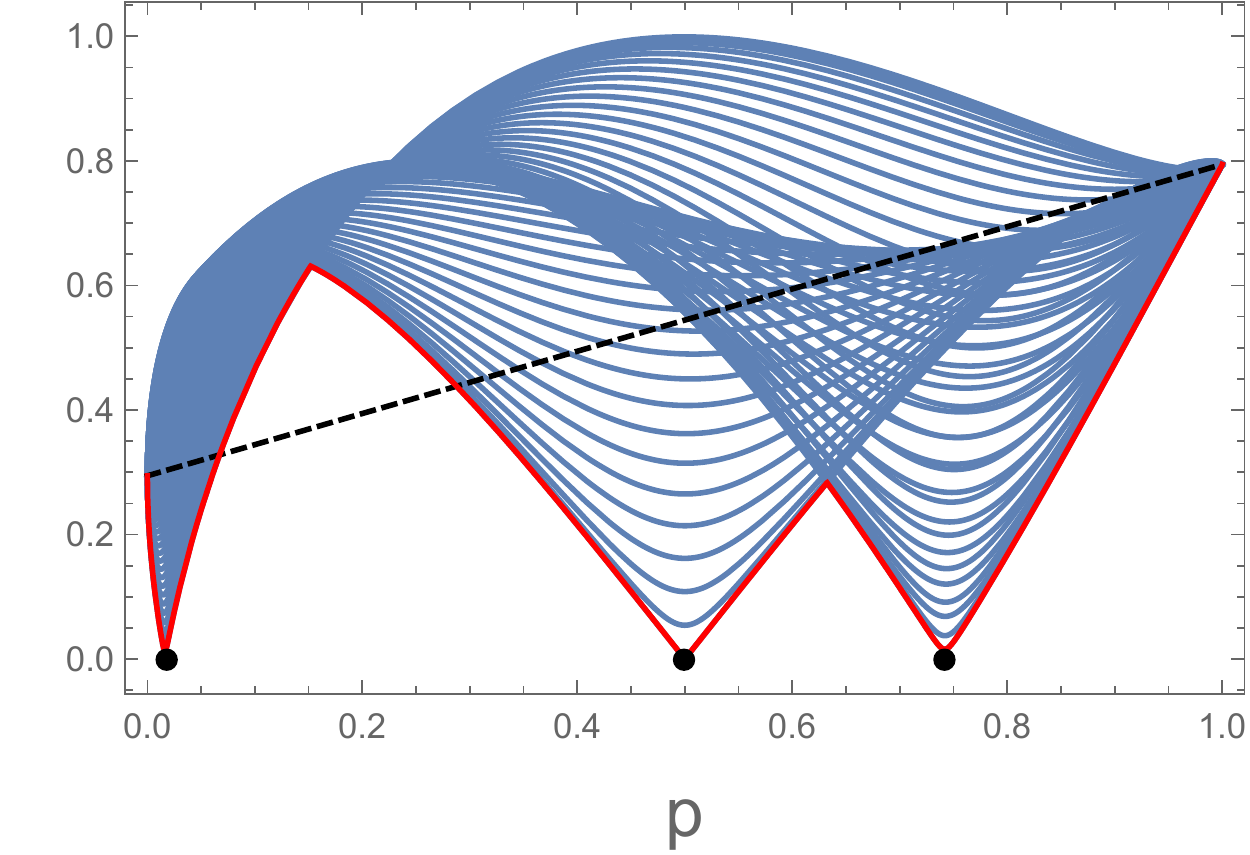}
\caption[fig5]{(Color online) (a) The $p$-dependence of $\tau_3 (p, 0)$, $\tau_3 (p, \pi)$, and $\tau_3 (p, \pi \pm \varphi_0)$ with $\varphi_0 = 1.27672$.
The dashed curve is a $p$-dependence of $\tau_3^{\max}$. The nontrivial zeros given in Table III are plotted as black dots. (b) The $p$-dependence of 
$\tau_3 (p, \varphi)$ with varying $\varphi$ from $0$ to $\pi$ with a step $0.05$. These curves have been referred as the characteristic curves. 
The red (lowest) solid line is a minimum of the characteristic curves. This red curve seems to indicate that $\tau_3 (\Pi)$ is zero at $p_1 \leq p \leq p_3$. However, we 
cannot find the corresponding optimal decompositions.}
\end{center}
\end{figure}

In Fig. 5(a) we plot $\tau_3 (p, \varphi)$ at $\varphi = 0$, $\pi \pm \varphi_0$, and $\pi$ with $\varphi_0 = 1.27642$. The nontrivial zeros 
$p_1$, $p_2$, and $p_3$ given at Table III are plotted as black dots. In Fig. 5(b) we plot $\tau_3 (p, \varphi)$ for various $\varphi$. 
These curves have been referred as the characteristic curves. The dashed line in both
figures is $p$-dependence of $\tau_3^{\max}$. The red (lowest) solid line in Fig. 5(b) is a minimum of the characteristic curves. The authors of Ref.\cite{oster07} have claimed that 
$\tau_3 (\Pi)$ is a convex hull of the minimum of the characteristic curves. If this is right, Fig. 5 (b) seems to exhibit that $\tau_3 (\Pi)$ is zero at 
$p_1 \leq p \leq p_3$. However, it is very difficult to find the corresponding 
optimal decompositions. For example, let us consider $p = p_3$ case. Table III indicates that the corresponding optimal decomposition is 
$(1 / 2) \left[ \ket{Z(p_3, \pi - \varphi_0)} \bra{Z(p_3, \pi - \varphi_0)} +  \ket{Z(p_3, \pi + \varphi_0)} \bra{Z(p_3, \pi + \varphi_0)} \right]$.
However, this is not equal to $\Pi (p_3)$ because of the cross terms. Similar difficulties arise at $p = p_1$ and $p = p_2$. So far, we do not know how to 
compute $\tau_3 (\Pi)$ analytically.

\end{appendix}


\begin{thebibliography}{99}
\bibitem{epr-35} A. Einstein, B. Podolsky and N. Rosen, {\it Can quantum-mechanical description of physical 
reality be considered complete ?}, Phys. Rev. {\bf A47} (1935) 777.
\bibitem{schrodinger-35} E. Schr\"{o}dinger, {\it Die gegenw\"{a}rtige Situation in der Quantenmechanik}, Naturwissenschaften, 
{\bf 23} (1935) 807.
\bibitem{text} M. A. Nielsen and I. L. Chuang, Quantum Computation and Quantum Information (Cambridge
University Press, Cambridge, England, 2000).
\bibitem{horodecki09} R. Horodecki, P. Horodecki, M. Horodecki, and K. Horodecki, {\it Quantum Entanglement}, Rev. Mod. Phys. 
{\bf 81} (2009) 865 [quant-ph/0702225] and references therein.
\bibitem{teleportation} C. H. Bennett, G. Brassard, C. Cr´epeau, R. Jozsa, A. Peres and W. K. Wootters, {\it Teleporting
an Unknown Quantum State via Dual Classical and Einstein-Podolsky-Rosen Channles}, Phys.Rev. Lett. {\bf 70} (1993) 1895.
\bibitem{superdense} C. H. Bennett and S. J. Wiesner, {\it Communication via one- and two-particle operators on
Einstein-Podolsky-Rosen states}, Phys. Rev. Lett. {\bf 69} (1992) 2881.
\bibitem{clon} V. Scarani, S. Lblisdir, N. Gisin and A. Acin, {\it Quantum cloning}, Rev. Mod. Phys. {\bf 77} (2005)
1225 [quant-ph/0511088] and references therein.
\bibitem{cryptography} A. K. Ekert , {\it Quantum Cryptography Based on Bell’s Theorem}, Phys. Rev. Lett. {\bf 67} (1991)
661.
\bibitem{cryptography2} C. Kollmitzer and M. Pivk, Applied Quantum Cryptography (Springer, Heidelberg, Germany, 2010).
\bibitem{qcreview} T. D. Ladd, F. Jelezko, R. Laflamme, Y. Nakamura, C. Monroe, and J. L. O'Brien, 
{\it Quantum Computers}, Nature, {\bf 464} (2010) 45 [arXiv:1009.2267 (quant-ph)].
\bibitem{computer} G. Vidal, {\it Efficient classical simulation of slightly entangled quantum computations}, Phys. Rev.
Lett. {\bf 91} (2003) 147902 [quant-ph/0301063].
\bibitem{benn96} C. H. Bennett, D. P. DiVincenzo, J. A. Smokin and W. K. Wootters,
{\it Mixed-state entanglement and quantum error correction}, Phys. Rev. {\bf A 54} (1996) 3824 [quant-ph/9604024].
\bibitem{vedral-97-1} V. Vedral, M. B. Plenio, M. A. Rippin and P. L. Knight, {\it Quantifying 
Entanglement}, Phys. Rev. Lett. {\bf 78} (1997) 2275 [quant-ph/9702027].
\bibitem{vedral-97-2} V. Vedral and M. B. Plenio, {\it Entanglement measures and purification procedures},
Phys. Rev. {\bf A 57} (1998) 1619 [quant-ph/9707035]. 
\bibitem{ree} A. Miranowicz and S. Ishizaka, {\it Closed formula for the relative entropy of entanglement}, 
Phys. Rev. {\bf A78} (2008) 032310 [arXiv:0805.3134 (quant-ph)]; H. Kim, M. R. Hwang, E. Jung and D. K. Park, {\it Difficulties in analytic computation for relative entropy of 
entanglement}, ibid. {\bf A 81} (2010) 052325 [arXiv:1002.4695 (quant-ph)]; D. K. Park, {Relative entropy of entanglement for two-qubit state with $z$-directional Bloch vectors}, 
Int. J. Quant. Inf. {\bf 8} (2010) 869 [arXiv:1005.4777 (quant-ph)]; S. Friedland and G Gour, {\it Closed formula for the relative entropy of entanglement in all dimensions}, 
J. Math. Phys. {\bf 52} (2011) 052201 [arXiv:1007.4544 (quant-ph)]; M. W. Girard, G.  Gour, and S. Friedland, {\it On convex optimization problems in quantum information theory}, 
arXiv:1402.0034 (quant-ph); E. Jung and D. K. Park, {\it REE From EOF}, Quant. Inf. Proc. {\bf 14} (2015) 531 arXiv:1404.7708 (quant-ph).
\bibitem{woot-98}S. Hill and W. K. Wootters, {\it Entanglement of a Pair of Quantum Bits}, Phys. Rev. Lett. {\bf 78} (1997) 5022 [quant-ph/9703041];
W. K. Wootters, {\it Entanglement of Formation of an Arbitrary State of Two Qubits}, ibid. {\bf 80} (1998) 2245 [quant-ph/9709029].
\bibitem{ckw} V. Coffman, J. Kundu and W. K. Wootters, {\it Distributed entanglement}, Phys. Rev. {\bf A 61} (2000) 052306 [quant-ph/9907047].
\bibitem{dur00} W. D\"{ur}, G. Vidal and J. I. Cirac, {\it Three qubits can be entangled in two inequivalent ways},
Phys.Rev. {\bf A 62} (2000) 062314 [quant-ph/0005115].
\bibitem{bennet00} C. H. Bennett, S. Popescu, D. Rohrlich, J. A. Smolin, and A. V. Thapliyal, {\it Exact and asymptotic measures
of multipartite pure-state entanglement}, Phys. Rev. {\bf A 63} (2000) 012307 [quant-ph/9908073].
\bibitem{uhlmann99-1} A. Uhlmann, {\it Fidelity and concurrence of conjugate states},
Phys. Rev. {\bf A 62} (2000) 032307 [quant-ph/9909060].
\bibitem{tangle} R. Lohmayer, A. Osterloh, J. Siewert and A. Uhlmann, {\it Entangled
Three-Qubit States without Concurrence and Three-Tangle}, Phys. Rev. Lett. {\bf 97}
(2006) 260502 [quant-ph/0606071];
C. Eltschka, A. Osterloh, J. Siewert and A. Uhlmann, {\it Three-tangle
for mixtures of generalized GHZ and generalized W states}, New J. Phys. {\bf 10} (2008)
043014 [arXiv:0711.4477 (quant-ph)];
E. Jung, M. R. Hwang, D. K. Park and J. W. Son, {\it Three-tangle
for Rank-$3$ Mixed States: Mixture of Greenberger-Horne-Zeilinger, W and flipped W states},
Phys. Rev. {\bf A 79} (2009) 024306 [arXiv:0810.5403 (quant-ph)];
E. Jung, D. K. Park, and J. W. Son, {\it Three-tangle does not properly
quantify tripartite entanglement for Greenberger-Horne-Zeilinger-type state},
Phys. Rev. {\bf A 80} (2009) 010301(R) [arXiv:0901.2620 (quant-ph)];
E. Jung, M. R. Hwang, D. K. Park, and S. Tamaryan, {\it Three-Party Entanglement in Tripartite Teleportation
Scheme through Noisy Channels}, Quant. Inf. Comp. {\bf 10} (2010) 0377 [arXiv:0904.2807 (quant-ph)].
\bibitem{elts12-1} C. Eltschka and J. Siewert, {\it Entanglement of Three-Qubit Greenberger-Horne-Zeilinger-Symmetric States}, 
Phys. Rev. Lett. {\bf 108} (2012) 020502 [ arXiv:1304.6095 (quant-ph)].
\bibitem{siewert12-1} J. Siewert and C. Eltschka, {\it Quantifying Tripartite Entanglement of Three-Qubit Generalized Werner States}, 
Phys. Rev. Lett. {\bf 108} (2012) 230502.
\bibitem{verst03} F. Verstraete, J. Dehaene, and D. De Moor, {\it Normal forms and entanglement measures for multipartite quantum states}, 
Phys. Rev. {\bf A 68} (2003) 012103 [quant-ph/0105090].
\bibitem{four-way} A. Osterloh and J. Siewert, {\it Constructing $N$-qubit entanglement monotones from antilinear operators},
Phys. Rev. {\bf A 72} (2005) 012337 [quant-ph/0410102]; D. \v{Z}. Dokovi\'{c} and A. Osterloh, {\it On polynomial invariants of several qubits},
J. Math. Phys. {\bf 50} (2009) 033509 [arXiv:0804.1661 (quant-ph)]; A. Osterloh and J. Siewert, 
{\it The invariant-comb approach and its relation to the balancedness of multiple entangled states}, 
New J. Phys. {\bf 12} (2010) 075025 [arXiv:0908.3818 (quant-ph)].
\bibitem{oster06-1} A. Osterloh and J. Siewert, {\it Entanglement monotones and maximally entangled states in multipartite qubit systems}, 
Quant. Inf. Comput. {\bf 4} (2006) 0531 [quant-ph/0506073].
\bibitem{eylee15-1} E. Jung and D. K. Park. {\it Entanglement of four-qubit rank-$2$ mixed states}, Quantum Inf. Process. {\bf 14} (2015) 3317
[arXiv:1505.06261  (quant-ph)].
\bibitem{osborne06-1} T. J. Osborne and F. Verstraete, {\it General Monogamy Inequality for Bipartite Qubit Entanglement}, Phys. Rev. Lett. {\bf 96}
(2006) 220503 [quant-ph/0502176].
\bibitem{bai07-1} Y.-K. Bai, D. Yang, and Z. D. Wang, {\it Multipartite quantum correlation and entanglement in four-qubit pure states}, Phys. Rev. 
{\bf A 76} (2007) 022336 [quant-ph/0703098].
\bibitem{bai08-1} Y.-K. Bai and Z. D. Wang. {\it Multipartite entanglement in four-qubit cluster-class states}, Phys. Rev. {\bf A 77} (2008) 032313
[arXiv:0709.4642 (quant-ph)].
\bibitem{regula14-1} B. Regula, S. D. Martino, S. Lee, and G. Adesso, {\it Strong Monogamy Conjecture for Multiqubit Entanglement: The Four-Qubit Case}, 
Phys. Rev. Lett. {\bf113} (2014) 110501 [arXiv:1405.3989 (quant-ph)].
\bibitem{jin15-1} J. H. Choi and J. S. Kim, {\it Negativity and strong monogamy of multiparty quantum entanglement beyond qubits}, 
Phys. Rev. {\bf A 92} (2015) 042307 [arXiv:1508.07673 (quant-ph)]
\bibitem{karmakar16-1} S. Karmakar, A. Sen, A. Bhar, and D. Sarkar, {\it Strong monogamy conjecture in a four-qubit system}, 
Phys. Rev. {\bf A 93} (2016) 012327 [arXiv:1512.06816 (quant-ph)]. 
\bibitem{vidal02} G. Vidal and R.F. Werner, {\it Computable measure of entanglement}, Phys. Rev. {\bf A 65} (2002) 032314 [quant-ph/0102117].
\bibitem{soojoon03} S. Lee, D. P. Chi, S. D. Oh, and J. Kim, {\it Convex-roof extended negativity as an entanglement measure for bipartite quantum systems},
Phys. Rev. {\bf A 68} (2003) 062304 [quant-ph/0310027].
\bibitem{oster07} A. Osterloh, J. Siewert and A. Uhlmann. {\it Tangles of superpositions and
the convex-roof extension}, Phys. Rev. {\bf A 77} (2008) 032310 [arXiv:0710.5909 (quant-ph)].








\end{thebibliography}
\end{document}